%% file: main.tex
\documentclass[sigconf]{acmart}
\usepackage{booktabs} 

\usepackage[english]{babel}
\usepackage{moresize}
\usepackage{amsmath}
\usepackage{algorithmic}
\usepackage{balance}
\usepackage{comment}
\usepackage{paralist}
\usepackage{bm}
\usepackage{pgfplots}
\usetikzlibrary{pgfplots.dateplot}

\usepackage{flushend}
\usepackage[english]{babel}
\usepackage[utf8]{inputenc}
\usepackage{mathrsfs}
\usepackage{svg}
\usepackage{graphicx}
\usepackage{xspace}
\usepackage{pifont}

\usepackage{threeparttable}

\usepackage{pifont}

\usepackage{amssymb}
\usepackage{amsfonts}
\usepackage{url}
\usepackage{longtable}
\usepackage{rotating}
\usepackage{multirow}
\usepackage{mathrsfs}
\usepackage{subfigure}
\usepackage{enumitem}
\usepackage[linesnumbered,algoruled,boxed,lined]{algorithm2e}
\usepackage{adjustbox}
\usepackage{hyperref}
\usepackage{pgfplots}
\usetikzlibrary{pgfplots.dateplot}
\usepackage{filecontents}
\usepackage{colortbl}

\newcommand{\Vodka}{\textsc{Winemaking}\xspace}

\newcommand{\hide}[1]{} 

\renewcommand\footnotetextcopyrightpermission[1]{}
\definecolor{bestcolor}{HTML}{FFCCC9}
\definecolor{secondbestcolor}{HTML}{FFFC9E}

\setcopyright{none}


\begin{document}
\fancyhead{}






\copyrightyear{2023}
\acmYear{2023}
\setcopyright{acmlicensed}\acmConference[WWW'23]{Proceedings of the ACM Web Conference 2023}{April 30-May 4, 2023}{Austin, TX, USA}
\acmBooktitle{Proceedings of the ACM Web Conference 2023 (WWW'23), April 30-May 4, 2023, Austin, TX, USA}
\acmPrice{15.00}
\acmDOI{10.1145/3543507.3583336}
\acmISBN{978-1-4503-9416-1/23/04}

\title{Winemaking: Extracting Essential Insights for Efficient Threat Detection in Audit Logs}

\author{Weiheng Wu}
\orcid{0009-0001-3114-7422}
\email{wuweiheng@iie.ac.cn}
\affiliation{%
	\institution{Institute of Information Engineering, Chinese Academy of Sciences}
	\institution{School of Cyber Security, University of Chinese Academy of Sciences}
	\city{Beijing}
	\country{China}
}

\author{Wei Qiao}
\email{qiaowei@iie.ac.cn}
\affiliation{%
	\institution{Institute of Information Engineering, Chinese Academy of Sciences}
	\institution{School of Cyber Security, University of Chinese Academy of Sciences}
	\city{Beijing}
	\country{China}
}

\author{Wenhao Yan}
\email{yanwenhao@iie.ac.cn}
\affiliation{%
	\institution{Institute of Information Engineering, Chinese Academy of Sciences}
	\institution{School of Cyber Security, University of Chinese Academy of Sciences}
	\city{Beijing}
	\country{China}
}

\author{Bo Jiang}
\email{jiangbo@iie.ac.cn}
\affiliation{%
	\institution{Institute of Information Engineering, Chinese Academy of Sciences}
	\institution{School of Cyber Security, University of Chinese Academy of Sciences}
	\city{Beijing}
	\country{China}
}

\author{Yuling Liu}
\email{liuyuling@iie.ac.cn}
\affiliation{%
	\institution{Institute of Information Engineering, Chinese Academy of Sciences}
	\institution{School of Cyber Security, University of Chinese Academy of Sciences}
	\city{Beijing}
	\country{China}
}

\author{Baoxu Liu}
\email{liubaoxu@iie.ac.cn}
\affiliation{%
	\institution{Institute of Information Engineering, Chinese Academy of Sciences}
	\institution{School of Cyber Security, University of Chinese Academy of Sciences}
	\city{Beijing}
	\country{China}
}

\author{Zhigang Lu}
\email{luzhigang@iie.ac.cn}
\affiliation{%
	\institution{Institute of Information Engineering, Chinese Academy of Sciences}
	\institution{School of Cyber Security, University of Chinese Academy of Sciences}
	\city{Beijing}
	\country{China}
}

\author{Junrong Liu}
\email{liujunrong@iie.ac.cn}
\affiliation{%
	\institution{Institute of Information Engineering, Chinese Academy of Sciences}
	\institution{School of Cyber Security, University of Chinese Academy of Sciences}
	\city{Beijing}
	\country{China}
}
\authornote{Corresponding author.}


\begin{abstract}
Advanced Persistent Threats (APTs) are continuously evolving, leveraging their stealthiness and persistence to put increasing pressure on current provenance-based Intrusion Detection Systems (IDS). This evolution exposes several critical issues: (1) The dense interaction between malicious and benign nodes within provenance graphs introduces neighbor noise, hindering effective detection; (2) The complex prediction mechanisms of existing APTs detection models lead to the insufficient utilization of prior knowledge embedded in the data; (3) The high computational cost makes detection impractical. 

To address these challenges, we propose \Vodka, a lightweight threat detection system built on a knowledge distillation framework, capable of node-level detection within audit log provenance graphs. Specifically, \Vodka applies graph Laplacian regularization to reduce neighbor noise, obtaining smoothed and denoised graph signals. Subsequently, \Vodka employs a teacher model based on GNNs to extract knowledge, which is then distilled into a lightweight student model. The student model is designed as a trainable combination of a feature transformation module and a personalized PageRank random walk label propagation module, with the former capturing feature knowledge and the latter learning label and structural knowledge. After distillation, the student model benefits from the knowledge of the teacher model to perform precise threat detection. Finally, \Vodka reconstructs attack paths from anomalous nodes, providing insight into the attackers' strategies. We evaluate \Vodka through extensive experiments on three public datasets and compare its performance against several state-of-the-art IDS solutions. The results demonstrate that \Vodka achieves outstanding detection accuracy across all scenarios 
and the detection time is 1.4 to 5.2 times faster than the current state-of-the-art methods. 

\end{abstract}


\maketitle
\input{intro}

\input{scene}
\input{solution}
\input{eval}

\input{relate}
\input{conclusion}


\clearpage
\bibliographystyle{ACM-Reference-Format}
\balance
\bibliography{reference}

\clearpage
 \input{appendix}

\end{document}

%% file: intro.tex
\section{Introduction}
\label{sec:intro}

Advanced Persistent Threats (APTs)\cite{alshamrani2019survey} represent a complex form of cyber attack characterized by high stealth and strong targeting. In these attacks, perpetrators gain unauthorized access to a victim's machine through methods such as network or software backdoors, and persist for extended periods to steal sensitive data or take control of the target machine. APTs have already infiltrated many highly secured large enterprises and institutions that are based on web services, causing substantial financial losses\cite{APT42, khaleefa2022concept, sharma2023advanced}. Notable examples include the Equifax\cite{Equifax} breach, which resulted in a record number of user data being stolen, and the SolarWinds\cite{Solarwinds} attack, which had a vast scope and severe impact.

To combat these sophisticated APTs attacks, host-based Intrusion Detection intrusion detection systems (IDS) have become a widely deployed defense mechanism. However, with the evolving nature of attack techniques and the growing scale of threats, traditional IDS solutions can no longer meet the demands of the changing threat landscape. Currently, provenance -based\cite{li2021threat} detection methods are considered effective for capturing APTs. These methods apply system audit logs and structure system entities (such as processes, files, and network flows) into a graph structure known as a provenance graph. Based on how audit logs are utilized, these detectors can be divided into three categories: statistics-based detection\cite{hassan2019nodoze, hassan2020we, liu2018towards} quantifies the suspiciousness of audit logs by analyzing the rarity in the provenance graph; rule-based detection\cite{hassan2020tactical, hossain2017sleuth, milajerdi2019holmes} matches audit logs with attack patterns using expert security knowledge bases; and learning-based detection\cite{jia2024magic, zengy2022shadewatcher, cheng2023kairos, wang2022threatrace, han2020unicorn, rehman2024flash} employs machine learning techniques to learn from the provenance graph, identifying abnormal system behaviors and attack patterns. Among these approaches, learning-based detection has been considered the most promising in recent years. However, we have observed several persistent challenges that impact real-world detection:

\begin{itemize}[leftmargin=*]
\item \textbf{Neighbor Noise}: In provenance graphs constructed from audit logs, malicious nodes often exhibit long-distance, multi-hop distributions, meaning that a complete attack can involve more benign system entities. The interaction between malicious and benign nodes generates interference and confounds the classifier in detection systems. For node-level classification tasks, the neighboring nodes adjacent to the target detection node introduce dense neighbor noise, which can obscure true anomalies, leading to false positives or missed detections, and ultimately reducing detection accuracy.
\item \textbf{High Computational Cost}: Graph-based algorithms enable learning from provenance graphs to capture the complex relationships within them. While these methods can achieve impressive detection performance, they come at the cost of significant memory and time overhead. As a result, these approaches lose the capability for real-time detection and can only function as offline systems to analyze graph data\cite{zengy2022shadewatcher, wang2022threatrace}.
\item \textbf{Insufficient Utilization of Prior Knowledge}: 
Recent APTs detection methods\cite{jia2024magic, zengy2022shadewatcher, cheng2023kairos} primarily rely on large-scale graph algorithms, such as GNNs, to extract structural and node feature information from graphs. However, the prediction mechanism of GNNs is highly complex, as it tightly integrates graph topology, node features, and projection matrices. This entanglement makes it challenging to clearly interpret the relationships between these factors within the model. This complexity prevents the effective utilization of prior knowledge in terms of labels, features, and structure\cite{li2019label, wang2020unifying}.
\end{itemize}


Thus, we propose \Vodka, a lightweight APTs detection method that achieves both high detection accuracy and computational efficiency. Specifically, \Vodka incorporates the following key functionalities:
\textbf{(1)} \Vodka constructs a provenance graph from audit logs that include various APTs attacks and uses Word2Vec to assign initial features to nodes. To address the issue of neighbor noise, \Vodka designs a graph Laplacian regularization-based graph signal denoising method tailored to the provenance graph. This algorithm smooths node signals without disrupting the original graph topology.
\textbf{(2)} \Vodka introduces a knowledge distillation framework where, during training, knowledge from a pre-trained large GNN teacher model is distilled into a lightweight student model. The lightweight student model is then used for anomalous node detection, significantly reducing detection costs.
\textbf{(3)} For the student model, \Vodka proposes a new hybrid mechanism that combines the Feature Transformation (FT) mechanism with a personalized PageRank random walk label propagation (PRL). This allows more effective learning of prior knowledge from features, labels, and structures while utilizing the GNN knowledge from the teacher model.
\textbf{(4)} After detecting anomalous nodes, \Vodka uses a community division algorithm to identify malicious communities.

We performed a comprehensive evaluation of \Vodka using datasets widely adopted by the research community, including StreamSpot\cite{Thestreamspotdataset}, Unicorn Wget\cite{han2020unicorn}, and DARPA-E3\cite{DARPA}. We first analyzed its detection performance and compared it against various existing baselines, demonstrating that \Vodka consistently outperforms almost all these systems under the same evaluation metrics. Furthermore, we analyzed \Vodka's detection costs and explored its potential as a real-time detection system. In addition, hyperparameter analysis and ablation experiments on \Vodka highlight the irreplaceability of its key parameters and main components. Lastly, we verified \Vodka's inherent robustness against adversarial attacks.
The primary contributions of our work are as follows:

\begin{figure}
	\centering
	\includegraphics[width=0.45\textwidth]{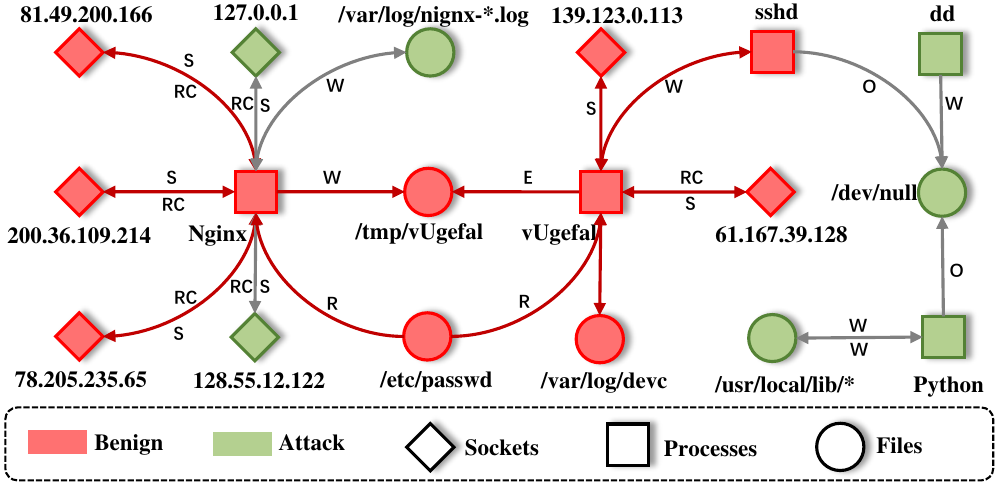}
	\caption{Attack scenario from DARPA E3 CADETS. Green indicates benign entities and red signifies malicious entities. R = Read, W = Write, O = Open, E = Execute, S = Send and Rc = Receive.}
	\Description{A figure illustrating four scenes.}
	\label{fig:scenario}
	\vspace{-0.2in}
\end{figure}

\begin{itemize}[leftmargin=*]

\item We propose \Vodka, a general APTs detection method based on knowledge distillation, capable of transferring GNN knowledge from a large teacher model to a lightweight student model, thereby reducing resource and time costs during detection.

\item We address the issue of neighbor noise in provenance graphs by designing a graph Laplacian regularization-based graph signal denoising method that smooths and denoises signals tailored for provenance graphs.

\item We examine the problem of insufficient utilization of prior knowledge caused by complex prediction mechanisms in current approaches and design a lightweight student model to effectively learn and leverage prior knowledge.

\item We comprehensively evaluate \Vodka’s performance by implementing it on three widely used datasets and conducting various experiments to verify its effectiveness. Experimental results demonstrate that \Vodka outperforms most existing methods while incurring lower detection costs.
\end{itemize}

%% file: scene.tex
\section{Attack scenarios and Threat model}
\label{sec:scene}

\subsection{Motivating Example}
This is an attack case from the DARPA E3 CADETS: where the attacker establishes a connection with the Nginx server via IP address 81.49.200.166 and gains shell access. Using the shell, the attacker downloads a malicious payload to /tmp/vUgefal and executes it. The process vUgefal then moves laterally to 61.167.39.128, completes the infection, and further injects sshd into the system, downloading another payload to /var/log/devc. Figure \ref{fig:scenario} illustrates our motivating example, where three types of nodes represent system entities, and the lines indicate the directional interactions between them.

\begin{table}[h]
	\centering
	\tiny
	\caption{Limitations of existing provenance-based intrusion detection systems.}
	\label{tab:limitation}
	\renewcommand\arraystretch{1.4}
	\resizebox{0.48\textwidth}{!}{
	\begin{tabular}{|c|c|c|c|c|c|}
		\hline
		\textbf{Model} & \textbf{\begin{tabular}[c]{@{}c@{}}Neighbor \\ Denoising\end{tabular}} & \textbf{\begin{tabular}[c]{@{}c@{}}Lightweight \\ Models\end{tabular}} & \textbf{\begin{tabular}[c]{@{}c@{}}Utilization of\\ Prior Knowledge\end{tabular}} & \textbf{\begin{tabular}[c]{@{}c@{}}Attack \\ Reconstruction\end{tabular}} & \textbf{Granularity} \\ \hline
		\textbf{\Vodka (ours)}     & \textcolor{green}{\ding{51}} & \textcolor{green}{\ding{51}} & \textcolor{green}{\ding{51}} & \textcolor{green}{\ding{51}} & Node \\ \hline
		\textbf{MAGIC\cite{jia2024magic}}     & \textcolor{green}{\ding{51}} & \textcolor{red}{\ding{55}}   & \textcolor{red}{\ding{55}} & \textcolor{red}{\ding{55}}   & Node \\ \hline
		\textbf{KAIROS\cite{cheng2023kairos}}    & \textcolor{green}{\ding{51}} & \textcolor{red}{\ding{55}}   & \textcolor{red}{\ding{55}}   & \textcolor{green}{\ding{51}} & Node \\ \hline
		\textbf{ThreaTrace\cite{wang2022threatrace}}& \textcolor{red}{\ding{55}}   & \textcolor{red}{\ding{55}}   & \textcolor{red}{\ding{55}}   & \textcolor{red}{\ding{55}}   & Node \\ \hline
		\textbf{ShadeWatcher\cite{zengy2022shadewatcher}} & \textcolor{red}{\ding{55}} & \textcolor{red}{\ding{55}}   & \textcolor{red}{\ding{55}}   & \textcolor{red}{\ding{55}}   & Edge \\ \hline
		\textbf{Unicorn\cite{han2020unicorn}}   & \textcolor{red}{\ding{55}}   & \textcolor{red}{\ding{55}}   & \textcolor{red}{\ding{55}}   & \textcolor{red}{\ding{55}}   & Graph \\ \hline
	\end{tabular}
}
\end{table}

\subsection{Limitations to Existing Solutions}
Learning-based detection methods are currently the most promising in the field of threat detection. Their core principle is to build anomaly models by learning from large datasets and then infer whether anomalous behavior represents a malicious attack. Some of the most notable recent works include:
\textbf{MAGIC\cite{jia2024magic}}, which performs efficient detection by masking the graph and reducing costs;
\textbf{KAIROS\cite{cheng2023kairos}}, which identifies four dimensions of intrusion detection systems and uses an encoder-decoder architecture to detect attacks and reconstruct attack chains;
\textbf{THREATRACE\cite{wang2022threatrace}}, the first detector to propose behavior pattern construction for each type of node in the provenance graph to achieve node-level detection;
\textbf{ShadeWatcherShadeWatcher\cite{zengy2022shadewatcher}}, which introduces recommendation system techniques into provenance graph detection and achieves edge-level detection;
\textbf{Unicorn\cite{han2020unicorn}}, which builds a provenance graph enriched with semantics and historical information to capture long-term stealthy attacks.
Although these works demonstrate excellent performance in threat detection, they still face limitations when tested in real-world environments. These limitations are what \Vodka aims to address, as shown in Table \ref{tab:limitation}.

\begin{itemize}[leftmargin=*]
	\item \textbf{Neighbor Denoising}:
	Provenance graphs constructed from audit logs contain rich information, with nodes and edges representing different meanings. However, current approaches seem too eager to move directly to graph representation learning without adequately addressing the noise in the initial graph. We observed that only Magic acknowledged this issue.
	
	\item \textbf{Lightweight Models}:
	One of the major challenge with current log provenance algorithms lies in the large size and difficulty of deployment of the models. Most existing approaches rely on GNN models, which, due to their multi-layer architectures and large parameter inputs, result in bloated models. Furthermore, the high memory and time overhead makes these detectors impractical and limits their deployment in real-world environments. Notably, only Magic\cite{jia2024magic} has optimized performance costs to address this issue.
	
	\item \textbf{Utilization of Prior Knowledge}:
	Labels, node features, and graph structures offer rich prior knowledge that can significantly enhance model performance. Label information helps the model capture the true class relationships of system entities. Node features reflect the attribute differences among system entities, and the graph structure reveals the topological relationships and latent patterns between interactions. However, existing APTs detection methods often over-rely on complex network topologies and automated feature learning, overlooking the explicit prior knowledge embedded in these components.
	
	\item \textbf{Attack Reconstruction}:
	A simple and complete attack chain helps security analysts better understand the attack behavior. However, from a large number of anomalous nodes detected at the node level, we need to reconstruct these attack paths based on the dependency relationships between kernel objects, as exemplified by KAIROS, to enable analysts to respond quickly.
	
	\item \textbf{Granularity}:
	Different detection granularities yield different levels of attack discovery. For example, graph-level granularity, as in Unicorn, is easy to partition but lacks the precision to pinpoint specific malicious entities in the graph. Edge-level granularity, as in ShadeWatcher, offers more detailed insights into specific edge relationships but incurs higher detection costs. Node-level granularity methods like Magic, KAIROS, and THREATRACE, can target specific entities without requiring detailed learning of edge relationships.
	
\end{itemize}

\subsection{Threat Model}

To build a more comprehensive threat model, we make several assumptions about attacker behavior and system characteristics. We consider APTs attackers, who aim to infiltrate a target host through network channels, pre-installed backdoors, or software vulnerabilities. This process is often highly stealthy and can persist for an extended period. During the intrusion, attackers attempt to blend their malicious activities with legitimate background data to obscure their intent. For instance, they may use Living-off-the-land techniques\cite{ongun2021living}, injecting malicious code into legitimate processes, which then spawn new legitimate processes to continue their malicious actions. 

Although the attackers' behavior is highly covert, audit logs capture complete, traceable evidence and footprints of their activities. As in most of the existing works in this field\cite{wang2022threatrace, han2020unicorn, rehman2024flash}, we assume that the underlying operating system, audit framework, and system hardware are intact and trustworthy\cite{bates2015trustworthy, pasquier2017practical}. Additionally, we assume that attackers cannot directly tamper with the contents of the audit logs, ensuring that the provenance graph constructed by \Vodka remains reliable\cite{paccagnella2020custos}.

%% file: solution.tex
\section{Methodology}
\label{sec:solution}

\begin{figure*}
	\centering
	\includesvg[width=1\textwidth]{figs/ov10_11.svg}
	\caption{The framework of \Vodka.}
	\label{fig:ov}
	\vspace{-0.1in}
\end{figure*}

In this section, we provide a detailed overview of \Vodka's overall design framework and the design of each module. The framework consists of the following components: \ding{172}\textbf{Graph Construction}. \ding{173}\textbf{Log Distillation}. \ding{174}\textbf{Threat Detection}. \ding{175}\textbf{Attack Reconstruction}. 
The overall framework is illustrated in Figure \ref{fig:ov}.

\subsection{Graph Construction}
\subsubsection{\bf Provenance Graph Construction}
\Vodka processes host audit logs from various sources(such as Windows ETW, Linux Audit and CamFlow\cite{pasquier2017practical}) and transforms them into a graph structure known as a Provenance Graph (PG). In this graph, nodes represent system entities, and edges denote the interactions between these entities. 


Since system logs contain rich attributes related to various system entities, \Vodka encodes these attributes into a vector space to serve as initial input for model learning. We use Word2Vec\cite{ma2015using}, a neural network-based technique effective at learning word vector representations, which generates dense, low-dimensional vectors for each word. In our cataloging process, we use attributes such as file paths for files, process names for process nodes, system entity types, and network IP addresses for sockets to generate feature vectors. These are combined with semantic attributes and system call types within one hop of the neighbors to form sentences. During training, \Vodka captures semantic relationships between words within a fixed length, outputting embedded vectors as the raw signal for subsequent modules.
 
\subsubsection{\bf Neighbor Denoising}
Given that in the provenance graph constructed from audit logs, malicious nodes (i.e., system entities manipulated by attackers) are rare, and adversarial activities—such as deceptive access to benign nodes or the manipulation of compromised benign nodes to perform misleading actions—are common, there exists a significant issue of neighbor noise for anomaly detection tasks. To address this, \Vodka incorporates a graph Laplacian regularization-based signal denoising method, which is designed to leverage the graph's topological structure to smooth signals and reduce the impact of noise.

Specifically, \Vodka defines the weights based on the similarity and connection relationships between entity nodes and combines them into a weight matrix $W$. By calculating the node degree matrix $D$, we can then derive the Laplacian matrix $L = D - W$. This enables the formulation of a convex quadratic optimization problem to minimize noise and smooth signals as follows:
\begin{equation}
	\min_{\mathbf{x}}\left\{\frac{1}{2}\|\mathbf{x}-\mathbf{x}^{(0)}\|^2+\frac{\gamma}{2}\mathbf{x}^\top\mathbf{L}\mathbf{x}\right\}.
\end{equation}
Where $||\cdot||$ represents the Euclidean norm, $x$ is the raw signal, and $\gamma$ is the regularization parameter that controls the trade-off between data fidelity and signal smoothing.
To achieve this, we solve the following linear system to find the optimal solution and obtain the denoised node signals:
\begin{equation}(\mathbf I+\gamma\mathbf L)\mathbf x=\mathbf x^{(0)},\quad\mathbf x=(\mathbf I+\gamma\mathbf L)^{-1}\mathbf x^{(0)}.\end{equation}

After the neighbor denoising process, we obtain denoised and smoothed signal vectors. It is worth noting that \Vodka's neighbor denoising algorithm does not alter the graph's topology in any way, ensuring that no new noise is introduced due to changes in the structure. This preserves the integrity of the original graph topology while effectively reducing the noise, allowing for more accurate and reliable detection in subsequent steps.

\subsection{Log Distillation}
In this section, we describe the knowledge distillation mechanism designed for \Vodka’s log provenance graph, which explains the training process of the teacher and student models and how knowledge is distilled from the large-scale teacher model to the lightweight student model.
\subsubsection{\bf Teacher Model}
The teacher model is the cornerstone of \Vodka’s APTs threat detection system, essential for reducing detection time costs. The teacher model is pre-trained before real-time detection occurs. For \Vodka’s teacher model, we employ various large-scale GNN models, such as GCN\cite{kipf2016semi}, GAT\cite{velivckovic2017graph}, GraphSAGE\cite{hamilton2017inductive}, and SGC\cite{wu2019simplifying}. These GNN models are well-suited for leveraging the complex interactions between system entities found in audit logs and capturing causal connections and semantic information across entities. This characteristic is commonly utilized by the most advanced IDS  today.

Unlike other IDS, \Vodka’s teacher model is trained using supervised labels on system entities fed into the GNN model, producing both the predicted result $f_thr$ and soft labels. Soft labels represent the aggregated anomaly probabilities for all test nodes and serve as an important self-supervised signal for the subsequent student model. By pre-training the teacher model on the large-scale graph topology derived from the log data and generating soft labels, \Vodka ensures robust support for real-time detection.

\subsubsection{\bf Student Model}
The core requirement for \Vodka’s student model is to compress detection time while maintaining high accuracy and effectively utilizing prior knowledge. Thus, the student model is designed to be lightweight and efficient, built on two foundational mechanisms: 1) feature transformation for individual nodes, and 2) label propagation\cite{iscen2019label} between nodes. The former focuses on enhancing the model’s ability to fit based on each node’s intrinsic features, while the latter propagates label information across nodes.

\Vodka’s student model integrates both mechanisms, combining \textbf{\underline{F}}eature \textbf{\underline{T}}ransformation \textbf{(FT)} and \textbf{\underline{P}}ersonalized PageRank \textbf{\underline{R}}andom walk \textbf{\underline{L}}abel propagation \textbf{(PRL)}. Feature Transformation enables each node to adjust and refine its feature representation to improve detection accuracy, while PRL helps spread labels effectively across the graph to capture anomaly information in a more efficient and informed manner. This hybrid approach allows the student model to leverage the knowledge distilled from the teacher model while significantly reducing detection time and computational costs.

\noindent\textbf{1. Feature Transformation}
We use a simple Multilayer Perceptron (MLP)\cite{tang2015extreme} for feature transformation to perform nonlinear transformations on the input node features, enhancing the model’s ability to fit complex data. In general, MLP takes the raw node features as input and outputs the predicted probability for malicious nodes. \Vodka employs a two-layer MLP.

Assuming the input feature vector for a node is $x_v$, with an input dimension of $d$, the MLP transforms it into a feature vector $h_v$ as follows:
\begin{equation}
	h_v=\sigma(W_1X_v+b_1).
\end{equation}
The prediction result from the FT module is 
\begin{equation}
	f_{\mathrm{FT}}(v)=\mathrm{softmax}(W_2h_v+b_2),
\end{equation}
where $W_1$ and $W_2$ are the learnable weight matrices corresponding to the first and second linear transformations, respectively. $b_1$ and $b_2$ are the bias terms, and $\sigma$ represents a non-linear activation function. The output layer passes through a softmax function to generate the final prediction.

\noindent\textbf{2.Label Propagation}
We designed a label propagation method based on Personalized PageRank\cite{bahmani2010fast} in the form of random walks\cite{xia2019random}. This approach propagates labels from labeled nodes to unlabeled nodes by simulating random walks between them. PRL introduces a “jump-back” mechanism, which allows the label information to propagate from neighbor nodes while also having a probability of returning to the starting node, preserving more local information.

The process can be broken down into the following steps:

\textbf{1) Label Initialization} For any node $v\in V$, its initial label is mapped to a probability distribution based on the supervised signal from the labeled data.

\textbf{2) Iterative Propagation} In each iteration, PPR updates the label distribution of each node based on its current distribution and that of its neighbors. Node $v$ retains its previous label distribution from the last iteration (denoted as iteration $k$) with a probability of $1 - \alpha$, meaning that it primarily relies on its own information without depending on its neighbors.

Then, with a probability $\alpha$, node $v$ receives label information from its neighboring nodes $N_v$. The label contribution from each neighbor $u\in{N_v}$ is an equally weighted average of the random walk label distribution $f_{RW}$. This ensures that each neighbor node contributes equally to the label propagation process, with all neighbors participating equally in updating the label distribution.

Finally, the updated label distribution $f_{\mathrm{RW}}^{k+1}(v)$ for node $v$ at iteration $k + 1$ is the sum of its retained distribution and the propagated distribution from its neighbors:
\begin{equation}
	f_{\mathrm{RW}}^{k+1}(v)=(1-\alpha)f_{\mathrm{RW}}^k(v)+\alpha\sum_{u\in N_v}\frac{f_{\mathrm{RW}}^k(u)}{|N_v|}.
\end{equation}
where $f_{\mathrm{RW}}^k(v)$ represents the label distribution of node $v$ during the $k$-th iteration, $\alpha$ is a parameter that controls the balance between jumping back and propagating the label, $N_v$ is the set of neighboring nodes of node $v$, and $|N_v|$ is the number of neighbors of node $v$, and $f_{\mathrm{RW}}^k(u)$ is the label distribution of node $u$.

\textbf{3) Stopping Iteration} The iteration stops when a fixed number of iterations is reached or when the label distribution converges. The result $f_{\mathrm{PRL}}^{k+1}(v) = f_{\mathrm{RW}}^{k+1}(v)$ is then returned.

\noindent\textbf{3. Combine tasks} After completing both FT and PRL, we combine them to form the complete student model task. During the learning process for each node $v$, we dynamically balance FT and PPR to make predictions. The prediction function is as follows:

\begin{equation}
	f_{std}^{k+1}(v)=\beta_{v}f_{FT}(v)+(1-\beta_{v})f_{\mathrm{PRL}}^{k+1}(v)
\end{equation}
Where $f_{FT}(v)$ is the prediction from the Feature Transformation module, $f_{\mathrm{PRL}}^{k+1}(v)$ is the result from the label propagation (PPR), and $\beta$ is a balancing parameter that controls the contribution of each component to the final prediction.

\subsubsection{Optimization Objective}
The training logic of both the teacher model and the student model has been presented. In summary, the teacher model distills knowledge obtained from large-scale GNN training, while the student model is trained to mimic the soft label predictions of the pre-trained teacher model. Thus, our optimization objective is to make the student model's prediction $f_{std}(v)$ approximate the teacher model's soft labels $f_{thr}(v)$. The optimization goal is expressed as:
\begin{equation}\min_{\Theta}\sum_{v\in V}distance(f_{thr}(v),f_{std;\Theta}(v))\end{equation}
Assuming that the student model performs $K$ iterations, the final optimization objective $F$ can be expressed as:
\begin{equation}\min_{\Theta}\sum_{\forall v\in V\}_{U}}\|f_{thr}(v)-f_{std;\Theta}^{K}(v)\|_{2}\end{equation}
In this context, $||\cdot||$ represents the L2 norm, and the parameter set $\theta$ includes the balancing parameters between PPR and FT, denoted as ${\beta_v, v\in{V}}$, the label jump-back and propagation balance parameter $\alpha$ in PPR, and the parameters $\theta$MLP from the MLP in the FT module.

\begin{figure}
	\centering
	\includegraphics[width=0.32\textwidth]{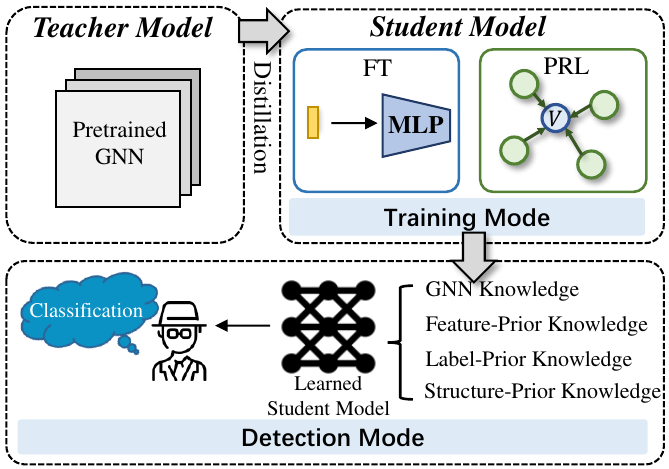}
	\caption{\Vodka working modes.}
	\Description{A figure illustrating four scenes.}
	\label{fig:design}
	\vspace{-0.1in}
\end{figure}

\subsection{Threat Detection}
 Our student model operates in two modes: a training mode and a detection mode. As shown in Figure \ref{fig:design}, during the knowledge distillation process, we transfer knowledge from the complex teacher model to the lightweight student model. In this phase, the student model is in training mode, focusing on learning the distilled knowledge from the teacher model. Once the knowledge distillation step is complete, \Vodka employs the trained student model as a detection model, which infers and classifies anomalous nodes from the provenance graph constructed from logs. At this point, the student model switches to detection mode.
The detection mode workflow is as follows: we input the provenance graph (with initial features) to the student model. After lightweight $K$-layer PRL and feature transformation steps, the student model predicts an anomaly score for each node $v$. The student model efficiently passes messages and aggregates information across the input graph, classifying nodes based on the learned lightweight feature representations. The anomaly score output for each node is then compared to a pre-set threshold, and nodes exceeding this threshold are flagged as potential malicious nodes and collected into the malicious node set.


\begin{table*}[]
	\centering
	\caption{Detection results with teacher models as GCN and GAT.}
	\vspace{-0.1in}
	\footnotesize
	\label{tab:GCN_GAT}
	\begin{tabular}{@{}c|c|cccc|c|c|cccc|c@{}}
		\toprule
		\multirow{2}{*}{Datasets} & \begin{tabular}[c]{@{}c@{}}Teacher\\ (GCN)\end{tabular} & \multicolumn{4}{c|}{Student}                                                                         & \multirow{2}{*}{\begin{tabular}[c]{@{}c@{}}+ACC\\ Impv.\end{tabular}} & \begin{tabular}[c]{@{}c@{}}Teacher\\ (GAT)\end{tabular} & \multicolumn{4}{c|}{Student}                                                                         & \multirow{2}{*}{\begin{tabular}[c]{@{}c@{}}+ACC\\ Impv.\end{tabular}} \\ \cmidrule(lr){2-6} \cmidrule(lr){8-12}
		& ACC                                                     & \multicolumn{1}{c|}{ACC}     & \multicolumn{1}{c|}{PR}      & \multicolumn{1}{c|}{RC}      & F1      &                                                                       & ACC                                                     & \multicolumn{1}{c|}{ACC}     & \multicolumn{1}{c|}{PR}      & \multicolumn{1}{c|}{RC}      & F1      &                                                                       \\ \midrule
		StreamSpot                & 98.30\%                                                 & \multicolumn{1}{c|}{99.57\%} & \multicolumn{1}{c|}{99.28\%} & \multicolumn{1}{c|}{99.81\%} & 99.54\% & 1.27\%                                                                & 97.93\%                                                 & \multicolumn{1}{c|}{99.82\%} & \multicolumn{1}{c|}{99.21\%} & \multicolumn{1}{c|}{99.35\%} & 99.28\% & 1.89\%                                                                \\ \midrule
		Unicorn Wget              & 98.21\%                                                 & \multicolumn{1}{c|}{99.16\%} & \multicolumn{1}{c|}{99.63\%} & \multicolumn{1}{c|}{99.76\%} & 99.69\% & 0.95\%                                                                & 98.27\%                                                 & \multicolumn{1}{c|}{99.75\%} & \multicolumn{1}{c|}{99.34\%} & \multicolumn{1}{c|}{99.82\%} & 99.58\% & 1.48\%                                                                \\ \midrule
		DARPA CADETS              & 97.31\%                                                 & \multicolumn{1}{c|}{99.77\%} & \multicolumn{1}{c|}{95.31}   & \multicolumn{1}{c|}{99.44\%} & 97.33\% & 2.46\%                                                                & 96.56\%                                                 & \multicolumn{1}{c|}{99.33\%} & \multicolumn{1}{c|}{95.53\%} & \multicolumn{1}{c|}{99.39\%} & 97.42\% & 2.77\%                                                                \\ \midrule
		DARPA TRACE               & 96.73\%                                                 & \multicolumn{1}{c|}{98.35\%} & \multicolumn{1}{c|}{95.73}   & \multicolumn{1}{c|}{99.02\%} & 97.34\% & 1.62\%                                                                & 96.22\%                                                 & \multicolumn{1}{c|}{98.98\%} & \multicolumn{1}{c|}{95.11\%} & \multicolumn{1}{c|}{99.40\%} & 97.21\% & 2.76\%                                                                \\ \midrule
		DARPA THEIA               & 97.26\%                                                 & \multicolumn{1}{c|}{99.94\%} & \multicolumn{1}{c|}{96.21}   & \multicolumn{1}{c|}{99.74\%} & 97.94\% & 2.68\%                                                                & 97.43\%                                                 & \multicolumn{1}{c|}{99.64\%} & \multicolumn{1}{c|}{96.40\%} & \multicolumn{1}{c|}{99.61\%} & 97.98\% & 2.21\%                                                                \\ \bottomrule
	\end{tabular}
\end{table*}

\subsection{Attack Reconstruction}
%
%
After threat detection, security personnel may have a set of detected malicious nodes provided by \Vodka, but these nodes are often dispersed across the graph, making it difficult to directly derive the attacker’s path. To alleviate the burden on security analysts, \Vodka reconstructs a more detailed attack trace, capturing the complete attack chain without relying on prior knowledge.
Specifically, \Vodka leverages Infomap\cite{edler2017infomap}, an algorithm based on random walks that simulates how information propagates between nodes and divides the graph into tightly connected communities. Malicious nodes and their one-hop neighbors are assigned higher weights to distinguish them from benign node connections. Next, information flow propagation is performed, where information spreads across the graph according to weighted probabilities, merging nodes to form communities. 

The goal is to maximize 'intra-community propagation' while minimizing 'inter-community propagation'. If most malicious nodes cluster within a single community, it indicates that the community may represent a core region of the APTs attack. If malicious nodes are distributed across multiple communities, it suggests that they likely serve as bridge nodes, linking different regions along the attack path.

%% file: eval.tex
\section{Evaluation}
\label{sec:eval}

In this section, we will conduct various experiments to validate the advantages of \Vodka and answer the following research questions:
\begin{itemize}[leftmargin=*]
\item \textbf{RQ1}: How does \Vodka’s detection performance compare to state-of-the-art methods?
\item \textbf{RQ2}: How effective are the main components of \Vodka?
\item \textbf{RQ3}: How do \Vodka’s hyperparameters affect its detection performance?
\item \textbf{RQ4}: What are the cost overheads of \Vodka as a detection system?
\item \textbf{RQ5}: How robust is \Vodka against adversarial attacks?
\end{itemize}

\subsection{\bf Datasets Settings}

\subsubsection{Datasets}\

\noindent\textbf{StreamSpot Dataset\cite{Thestreamspotdataset}}: This dataset comprises provenance graphs collected by SystemTap from six controlled environments, including CNN, Download, Gmail, Vgame, YouTube, and Attack, with each environment containing 100 provenance graphs.

\noindent\textbf{Unicorn Wget Dataset\cite{han2020unicorn}}: This dataset includes simulated attacks designed by UNICORN, consisting of 150 batches of log data collected by CamFlow, of which 125 batches are benign and 25 are malicious.

\noindent\textbf{DARPA-E3 Dataset\cite{DARPA}}: This dataset is part of the DARPA TC program's third evaluation, containing several sub-datasets. We selected the Cadets, Theia, and Trace datasets to evaluate \Vodka, using the same ground truth labels as ThreaTrace. Detailed descriptions of these datasets can be found in the appendix.

\subsubsection{Metrics}
We use the evaluation metrics from Unicorn\cite{han2020unicorn} and ThreaTrace\cite{wang2022threatrace} to compare performance across different granularities, enabling reasonable comparisons with other detection methods. The performance metrics include Precision (PR), Recall (RC), F1-score and Accuracy (ACC).

\subsubsection{Baselines}
To perform a comprehensive performance evaluation, we compared \Vodka with the best detection methods. Unfortunately, some methods were excluded due to challenges in reproducibility or incompatibility. For example,  KAIROS\cite{cheng2023kairos} could not be compared as it uses a time window-based approach that differs from other methods. We selected the following detection methods as our baselines:

\begin{itemize}[leftmargin=*]
\item \textbf{StreamSpot\cite{jacob2008systemtap}}: StreamSpot detects intrusions by analyzing information flow graphs. It extracts features from the graph to learn a benign model and uses clustering methods to detect anomalous graphs.
\item \textbf{Unicorn\cite{han2020unicorn}}: Unicorn utilizes graph rendering techniques to efficiently summarize long-running system executions. It classifies graphs as benign or malicious and filters anomalies.
\item \textbf{Prov-Gem\cite{kapoor2021prov}}: Prov-Gem captures entity interactions using a unified relational-aware embedding framework, with the help of supervised signals.
\item \textbf{ThreaTrace\cite{wang2022threatrace}}: ThreaTrace customizes a new GraphSAGE method to aggregate system entity nodes, enabling node-level detection.
\item \textbf{Log2vec\cite{liu2019log2vec}}: Log2vec is a heterogeneous graph embedding-based method for detecting network threats within enterprises. It identifies abnormal logs using node embeddings and clustering methods.
\item \textbf{DeepLog\cite{du2017deeplog}}: DeepLog leverages Long Short-Term Memory (LSTM)\cite{yu2019review} networks to model system logs as natural language sequences and performs log-level anomaly detection.
\item \textbf{MAGIC\cite{jia2024magic}}: MAGIC is a multi-granularity detection method that achieves efficient detection and cost reduction through graph masking techniques.
\end{itemize}

\subsubsection{Implementation}
 Details We implemented \Vodka’s prototype in approximately 2800 lines of Python3.11 code. For log processing, we used a log parser to convert audit log files from various sources into JSON format and then preprocess them. The development environment was PyTorch\cite{paszke2019pytorch}, and the model implementation was supported by the Deep Graph Library (DGL)\cite{wang2019deep}. Detailed teacher and student model parameters are provided in the \ref{sec:appendix} Appendix.

\begin{table*}[]
	\centering
	\caption{Detection results with teacher models as SAGE and SGC.}
	\vspace{-0.1in}
	\footnotesize
	\label{tab:SAGE_SGC}
	\begin{tabular}{@{}c|c|cccc|c|c|cccc|c@{}}
		\toprule
		\multirow{2}{*}{Datasets} & \begin{tabular}[c]{@{}c@{}}Teacher\\ (SAGE)\end{tabular} & \multicolumn{4}{c|}{Student}                                                                         & \multirow{2}{*}{\begin{tabular}[c]{@{}c@{}}+ACC\\ Impv.\end{tabular}} & \begin{tabular}[c]{@{}c@{}}Teacher\\ (SGC)\end{tabular} & \multicolumn{4}{c|}{Student}                                                                         & \multirow{2}{*}{\begin{tabular}[c]{@{}c@{}}+ACC\\ Impv.\end{tabular}} \\ \cmidrule(lr){2-6} \cmidrule(lr){8-12}
		& ACC                                                      & \multicolumn{1}{c|}{ACC}     & \multicolumn{1}{c|}{PR}      & \multicolumn{1}{c|}{RC}      & F1      &                                                                       & ACC                                                     & \multicolumn{1}{c|}{ACC}     & \multicolumn{1}{c|}{PR}      & \multicolumn{1}{c|}{RC}      & F1      &                                                                       \\ \midrule
		StreamSpot                & 98.10\%                                                  & \multicolumn{1}{c|}{99.00\%} & \multicolumn{1}{c|}{99.13\%} & \multicolumn{1}{c|}{99.27\%} & 99.20\% & 0.90\%                                                                & 98.24\%                                                 & \multicolumn{1}{c|}{99.77\%} & \multicolumn{1}{c|}{99.47\%} & \multicolumn{1}{c|}{99.26\%} & 99.36\% & 1.53\%                                                                \\ \midrule
		Unicorn Wget              & 97.34\%                                                  & \multicolumn{1}{c|}{99.34\%} & \multicolumn{1}{c|}{99.23\%} & \multicolumn{1}{c|}{99.15\%} & 99.19\% & 2.00\%                                                                & 98.00\%                                                 & \multicolumn{1}{c|}{99.36\%} & \multicolumn{1}{c|}{99.66\%} & \multicolumn{1}{c|}{99.50\%} & 99.58\% & 1.36\%                                                                \\ \midrule
		DARPA CADETS              & 97.20\%                                                  & \multicolumn{1}{c|}{99.03\%} & \multicolumn{1}{c|}{94.87\%} & \multicolumn{1}{c|}{99.78\%} & 97.26\% & 1.83\%                                                                & 97.03\%                                                 & \multicolumn{1}{c|}{99.15\%} & \multicolumn{1}{c|}{94.13\%} & \multicolumn{1}{c|}{99.71\%} & 96.84\% & 2.12\%                                                                \\ \midrule
		DARPA TRACE               & 96.03\%                                                  & \multicolumn{1}{c|}{97.45\%} & \multicolumn{1}{c|}{95.22\%} & \multicolumn{1}{c|}{99.34\%} & 97.23\% & 1.42\%                                                                & 96.18\%                                                 & \multicolumn{1}{c|}{97.60\%} & \multicolumn{1}{c|}{93.28\%} & \multicolumn{1}{c|}{99.35\%} & 96.24\% & 1.42\%                                                                \\ \midrule
		DARPA THEIA               & 96.93\%                                                  & \multicolumn{1}{c|}{98.94\%} & \multicolumn{1}{c|}{95.94\%} & \multicolumn{1}{c|}{99.40\%} & 97.65\% & 2.01\%                                                                & 97.29\%                                                 & \multicolumn{1}{c|}{99.23\%} & \multicolumn{1}{c|}{94.70\%} & \multicolumn{1}{c|}{99.22\%} & 96.94\% & 1.94\%                                                                \\ \bottomrule
	\end{tabular}
\end{table*}

\subsection{Overall Detection Efficacy Comparsion(Q1)}
Table \ref{tab:GCN_GAT} and Table \ref{tab:SAGE_SGC} show \Vodka’s performance across various datasets using different teacher models. We observe that \Vodka’s student model consistently improves ACC after distillation from all teacher models. Table \ref{tab:comparison} compares \Vodka with the baselines across multiple datasets. For StreamSpot, we compare \Vodka with StreamSpot, ThreaTrace, and MAGIC; for Unicorn Wget, we compare \Vodka with Unicorn, Prov-Gem, ThreaTrace, and MAGIC; for DARPA E3 datasets, we compare \Vodka with Prov-Gem, ThreaTrace, Log2vec, DeepLog, and MAGIC. Results indicate that \Vodka achieves the best or second-best results on all datasets.
Specifically, on the \textbf{StreamSpot}, which involves relatively simple attack scenario graphs, \Vodka—being an entity-level detector—handles it quite easily. In the \textbf{Unicorn Wget}, which is somewhat more complex compared to StreamSpot but still graph-level, \Vodka only slightly lags behind Prov-Gem in Precision (PR), but outperforms all other baselines across all other metrics. In the three \textbf{DARPA E3}, \Vodka ranks second only to MAGIC in the Theia and Trace datasets, slightly behind in PR and F1-score.

These results demonstrate that \Vodka’s knowledge distillation approach effectively transfers knowledge from complex GNN teacher models to the student model, enabling the student model to retain high anomaly detection accuracy across multiple detection mechanisms.

\begin{table}[]
	\centering
	\caption{\Vodka’ comparison results.}
	\label{tab:comparison}
	\vspace{-0.1in}
	\footnotesize
	\renewcommand\arraystretch{1.2}
	\begin{threeparttable}
		\begin{tabular}{c|c|cccc}
			\hline
			Datasets                                                                  & Systems    & ACC                                    & PR                                     & RC                                     & F1                                    \\ \hline
			& StreamSpot & 66\%                                   & 73\%                                   & 91\%                                   & 81\%                                  \\ \cline{2-2}
			& ThreaTrace & 96\%                                   & 95\%                                   & 93\%                                   & 96\%                                  \\ \cline{2-2}
			& MAGIC      & \cellcolor[HTML]{FFFC9E}99\%           & \cellcolor[HTML]{FFFC9E}99\%           & \cellcolor[HTML]{FFFC9E}99\%           & \cellcolor[HTML]{FFFC9E}99\%          \\ \cline{2-2}
			\multirow{-4}{*}{StreamSpot}                                              & \Vodka      & \cellcolor[HTML]{FFCCC9}\textbf{99\%}  & \cellcolor[HTML]{FFCCC9}\textbf{99\%}  & \cellcolor[HTML]{FFCCC9}\textbf{100\%} & \cellcolor[HTML]{FFCCC9}\textbf{99\%} \\ \hline
			& Unicorn    & 90\%                                   & 86\%                                   & 95\%                                   & 90\%                                  \\ \cline{2-2}
			& Prov-Gem   & NA                                     & \cellcolor[HTML]{FFCCC9}\textbf{100\%} & 80\%                                   & 89\%                                  \\ \cline{2-2}
			& ThreaTrace & 98\%                                   & 93\%                                   & \cellcolor[HTML]{FFFC9E}98\%           & 95\%                                  \\ \cline{2-2}
			& MAGIC      & \cellcolor[HTML]{FFFC9E}99\%           & 98\%                                   & 96\%                                   & \cellcolor[HTML]{FFFC9E}97\%          \\ \cline{2-2}
			\multirow{-5}{*}{\begin{tabular}[c]{@{}c@{}}Unicorn\\ Wget\end{tabular}}  & \Vodka      & \cellcolor[HTML]{FFCCC9}\textbf{99\%}  & \cellcolor[HTML]{FFFC9E}99\%           & \cellcolor[HTML]{FFCCC9}\textbf{99\%}  & \cellcolor[HTML]{FFCCC9}\textbf{99\%} \\ \hline
			& Log2vec    & 98\%                                   & 49\%                                   & 85\%                                   & 62\%                                  \\ \cline{2-2}
			& DeepLog    & 95\%                                   & 23\%                                   & 74\%                                   & 35\%                                  \\ \cline{2-2}
			& ThreaTrace & \cellcolor[HTML]{FFFC9E}99\%           & 90\%                                   & \cellcolor[HTML]{FFFC9E}99\%           & 95\%                                  \\ \cline{2-2}
			& MAGIC      & \cellcolor[HTML]{FFFC9E}99\%           & \cellcolor[HTML]{FFFC9E}94\%           & \cellcolor[HTML]{FFFC9E}99\%           & \cellcolor[HTML]{FFFC9E}97\%          \\ \cline{2-2}
			\multirow{-5}{*}{\begin{tabular}[c]{@{}c@{}}DARPA \\ CADETS\end{tabular}} & \Vodka      & \cellcolor[HTML]{FFCCC9}\textbf{99\%}  & \cellcolor[HTML]{FFCCC9}\textbf{95\%}  & \cellcolor[HTML]{FFCCC9}\textbf{99\%}  & \cellcolor[HTML]{FFCCC9}\textbf{97\%} \\ \hline
			& Log2vec    & 97\%                                   & 54\%                                   & 78\%                                   & 64\%                                  \\ \cline{2-2}
			& DeepLog    & 96\%                                   & 41\%                                   & 68\%                                   & 51\%                                  \\ \cline{2-2}
			& ThreaTrace & \cellcolor[HTML]{FFFC9E}98\%           & 72\%                                   & \cellcolor[HTML]{FFFC9E}99\%           & 83\%                                  \\ \cline{2-2}
			& MAGIC      & \cellcolor[HTML]{FFCCC9}\textbf{99\%}  & \cellcolor[HTML]{FFCCC9}\textbf{99\%}  & \cellcolor[HTML]{FFFC9E}99\%           & \cellcolor[HTML]{FFCCC9}\textbf{99\%} \\ \cline{2-2}
			\multirow{-5}{*}{\begin{tabular}[c]{@{}c@{}}DARPA \\ TRACE\end{tabular}}  & \Vodka      & \cellcolor[HTML]{FFFC9E}98\%           & \cellcolor[HTML]{FFFC9E}95\%           & \cellcolor[HTML]{FFCCC9}\textbf{99\%}  & \cellcolor[HTML]{FFFC9E}97\%          \\ \hline
			& Log2vec    & 99\%                                   & 62\%                                   & 66\%                                   & 64\%                                  \\ \cline{2-2}
			& DeepLog    & 98\%                                   & 16\%                                   & 14\%                                   & 15\%                                  \\ \cline{2-2}
			& ThreaTrace & \cellcolor[HTML]{FFFC9E}99\%           & 87\%                                   & \cellcolor[HTML]{FFFC9E}99\%           & 93\%                                  \\ \cline{2-2}
			& MAGIC      & \cellcolor[HTML]{FFFC9E}99\%           & \cellcolor[HTML]{FFCCC9}\textbf{98\%}  & \cellcolor[HTML]{FFFC9E}99\%           & \cellcolor[HTML]{FFFC9E}98\%          \\ \cline{2-2}
			\multirow{-5}{*}{\begin{tabular}[c]{@{}c@{}}DARPA \\ THEIA\end{tabular}}  & \Vodka      & \cellcolor[HTML]{FFCCC9}\textbf{100\%} & \cellcolor[HTML]{FFFC9E}96\%           & \cellcolor[HTML]{FFCCC9}\textbf{99\%}  & \cellcolor[HTML]{FFCCC9}\textbf{98\%} \\ \hline
		\end{tabular}
		
		\begin{tablenotes}
			\item - \raisebox{0.5ex}{\fcolorbox{black}{bestcolor}{\rule{0pt}{1pt}\rule{6pt}{0pt}}} Best performance
			\item - \raisebox{0.5ex}{\fcolorbox{black}{secondbestcolor}{\rule{0pt}{1pt}\rule{6pt}{0pt}}} Second best performance
		\end{tablenotes}

		\vspace{-0.1in}
		
	\end{threeparttable}
\end{table}

\vspace{-0.1in}

\subsection{Ablation Study (RQ2)}
\begin{figure}[t]
	\centering
	\subfigure[FT]{
		\includegraphics[width=0.45\columnwidth]{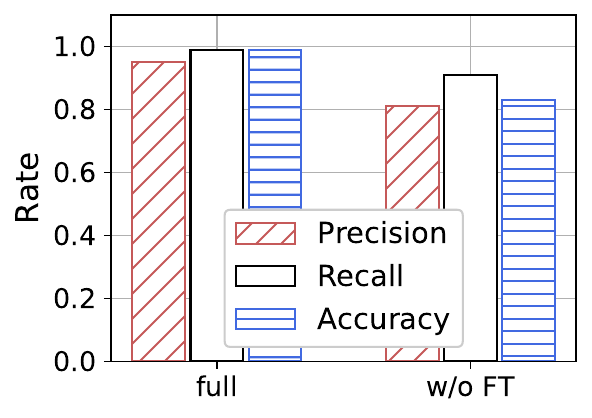}
	}
	\subfigure[PRL]{
		\includegraphics[width=0.45\columnwidth]{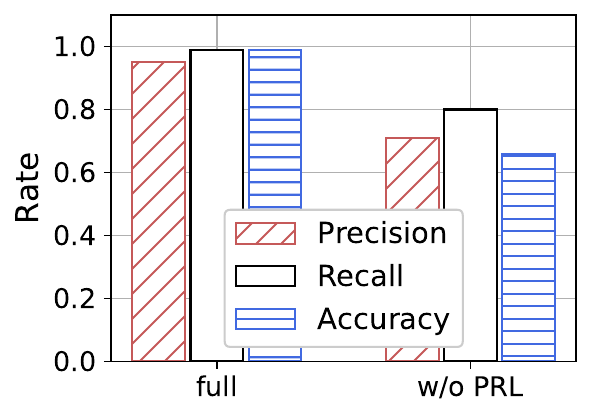}
	}
	\vspace{-0.1in}
	\caption{Ablation study}
	\label{fig:ablation}
	\vspace{-0.1in}
\end{figure}

We conducted an ablation study on the CADETS dataset using GCN as the teacher model to investigate the individual contributions of \Vodka’s different submodules, as shown in Figure \ref{fig:ablation}.

\noindent\textbf{Impact of Feature Transformation}: We removed the feature transformation mechanism from the student model, leaving only the label propagation mechanism. As shown in Figure \ref{fig:ablation}.(a), removing the feature transformation significantly reduced performance—PR decreased by 14$\%$,  RC by 8$\%$, and ACC by 16$\%$,. This demonstrates that the feature-based prior knowledge is crucial for deep learning in node inference tasks. Without it, the student model loses the ability to learn deeply from system entity attributes.

\noindent\textbf{Impact of Label Propagation}: We removed the label propagation mechanism (PRL) from the student model, leaving only the feature transformation mechanism. As shown in Figure \ref{fig:ablation}.(b), the absence of PRL led to an even larger performance drop—PR decreased by 24$\%$, RC by 19$\%$, and ACC by 33$\%$. This highlights that PRL’s contribution of structural and label-based prior knowledge is significant, enabling the student model to leverage contextual and temporal information effectively.

\begin{figure}[t]
	\centering
	\subfigure[Embedding Dimension]{
		\includegraphics[width=0.45\columnwidth]{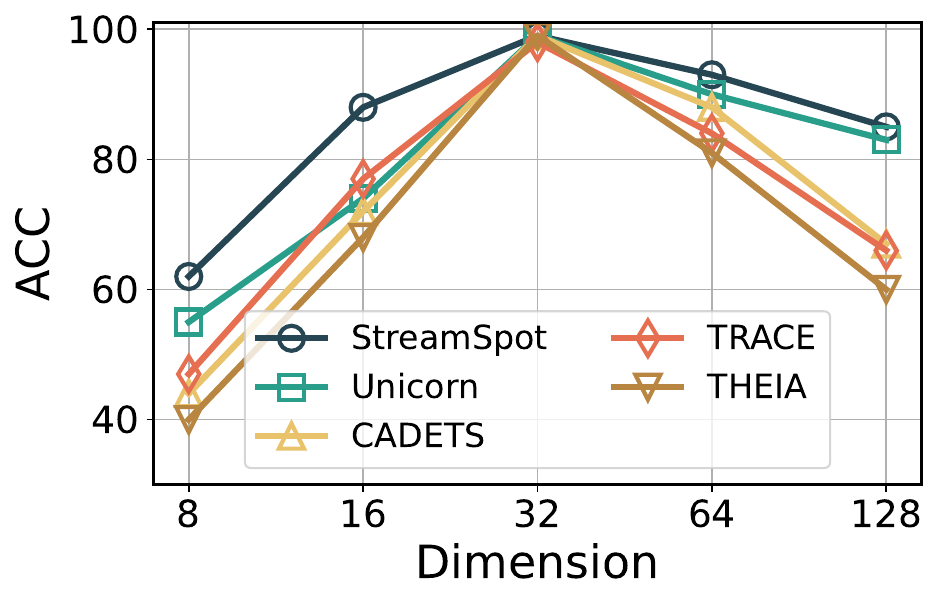}
	}
	\subfigure[Training Ratio]{
		\includegraphics[width=0.45\columnwidth]{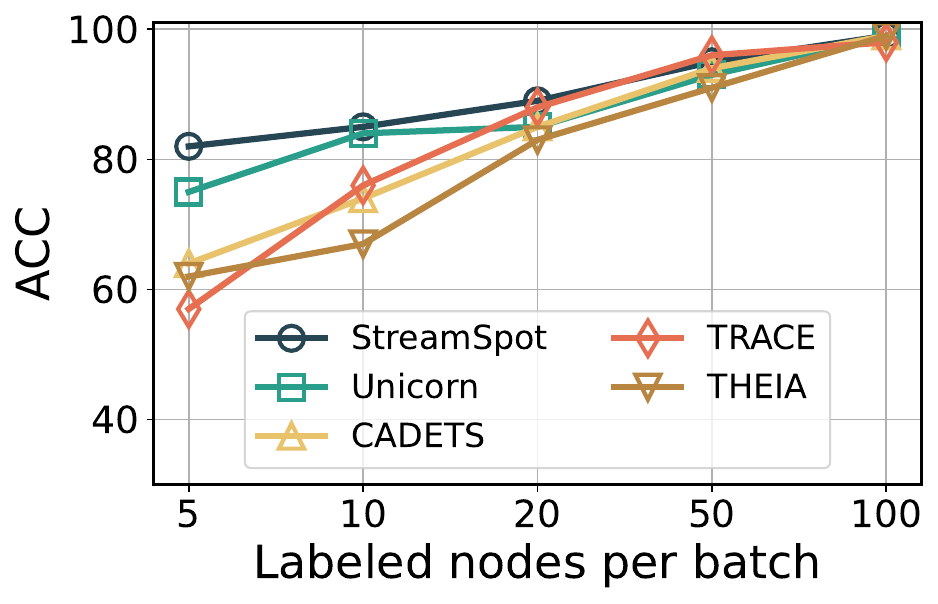}
	}
	\vspace{-0.1in}
	\caption{Hyperparameter Investigation}
	\label{fig:hy}
	\vspace{-0.1in}
\end{figure}

\subsection{Hyperparameter Investigation (RQ3)}
We conducted the following experiments to explore how key hyperparameters affect ACC. Using GCN as the teacher model, we tested the entire dataset.


\noindent\textbf{Embedding Dimension}: The embedding dimension determines the length of the vector representing each node. We tested dimensions {8, 16, 32, 64, 128}. Figure \ref{fig:hy}.(a) shows that lower dimensions result in poorer performance due to limited expressiveness. The optimal performance was achieved at 32 dimensions. Higher dimensions did not improve performance and even slightly decreased it due to redundant features, increasing overfitting risk.


\noindent\textbf{Training Ratio}: The training ratio determines the number of labeled nodes in each batch. To further demonstrate the effectiveness of Vodka, we tested values of {5, 10, 20, 50, 100}. The experimental results are shown in Figure \ref{fig:hy}.(b). As we can observe, the classification performance of the student model improves as the number of labeled nodes increases, reaching near-optimal performance when the batch contains 100 labeled nodes.
We attempt to analyze the reasons behind this phenomenon: Increasing the number of labeled nodes introduces more supervised signals, which in turn enhances the efficiency of label inference during PRL. For a semi-supervised system like \Vodka, this improvement in label propagation is undoubtedly significant in boosting the model's performance.

\subsection{Performance Overhead (RQ4)}
\Vodka aims to perform APTs detection with minimal overhead and flexible deployment in various environments. In real-world deployments, GPUs may be unavailable, and only CPUs may be used. Therefore, we primarily focused on \Vodka’s detection time and memory usage under CPU-only conditions. Table \ref{tab:time} compares \Vodka’s performance overhead with ShadeWatcher and MAGIC on the DARPA E3 TRACE dataset. Results show that \Vodka has significant advantages in both detection time and memory usage. At the same training ratio, \Vodka is 1.4 times faster than ShadeWatcher and 5.2 times faster than MAGIC.

Given its high detection speed, \Vodka could be deployed as a real-time online detector. For example, with the TRACE dataset, which took two weeks to collect, generating 1.37GB of audit logs per day, \Vodka only requires 23 seconds per day to complete the APTs detection. This means \Vodka’s lightweight and high-speed detection capabilities make it suitable for daily APTs monitoring in small organizations and enterprises.

\begin{figure}[t]
	\centering
	\includegraphics[width=0.9\columnwidth]{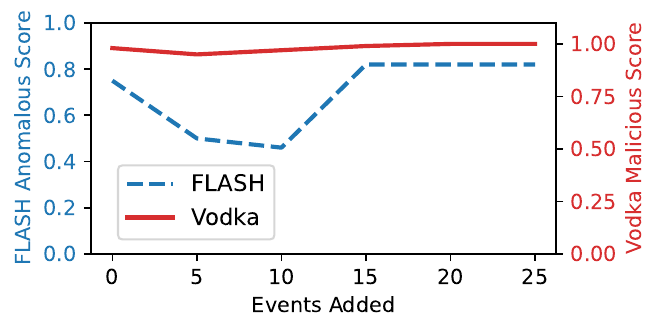}
	\vspace{-0.08in}
	\caption{Robustness against adersarial attacks study.}
	\vspace{-0.1in}
	\label{fig:adver}
	\Description{A line figure showing the performance change with respect to the number of centric nodes, the masking subgraph scale, and the self-contrastive weight.}
	\vspace{-0.12in}
\end{figure}

\subsection{Robustness Against Adversarial Attacks (RQ5)}
As more works focus on provenance-based data analysis, adversarial attacks\cite{chakraborty2021survey, tramer2019adversarial} designed to evade such provenance-based detectors are also increasing. One common adversarial attack is adversarial mimicry, which alters provenance data and adds benign representations to mimic benign behavior, disguising malicious system entities to evade detection. In this section, we simulate mimicry attacks to evaluate \Vodka’s robustness against adversarial mimicry, using evaluation standards from \cite{goyal2023sometimes} and \cite{rehman2024flash}.
Our mimicry attack steps are as follows: first, we select benign system entities from the normal training data. Next, we introduce events and combine them with the benign entities. Finally, we embed the structure of the benign nodes into the attack graph. We compare \Vodka (using GCN as the teacher model and CADETS as the dataset) with FLASH\cite{rehman2024flash}. Results (Figure \ref{fig:adver}) show that, as false events increase, FLASH’s anomaly score significantly decreases, indicating its weak robustness against mimicry attacks. In contrast, \Vodka maintains an anomaly score above 95\% for all levels off false events, demonstrating its superior robustness.

The reasons for this are twofold: first, \Vodka is based on entity-level detection, making it highly sensitive to large-scale false event additions. Second, the provenance graph’s neighbor denoising process smooths out some of the strong malicious expressions, making it less susceptible to manipulation.

\begin{table}[]
	\centering
	\caption{\Vodka’ performance overhead.}
	\vspace{-0.1in}
	\label{tab:time}
	\footnotesize
	\renewcommand\arraystretch{1.2}
	\begin{tabular}{c|c|c|c}
		\hline
		System       & Granularity & \begin{tabular}[c]{@{}c@{}}Time \\ consumpution(s)\end{tabular} & \begin{tabular}[c]{@{}c@{}}Peak Memory \\ consumption (MB)\end{tabular} \\ \hline
		ShadeWatcher & Edge        & 220                                                             & NA                                                                      \\ \hline
		MAGIC        & Node        & 825                                                             & 1,667                                                                   \\ \hline
		\Vodka        & Node        & \textbf{158}                                                             & \textbf{982}                                                                     \\ \hline
	\end{tabular}
	\vspace{-0.13in}
\end{table}

%% file: relate.tex
\section{Related Work}
\label{sec:relate}

\noindent \textbf{APTs Detection}. 
Current research primarily focuses on provenance-based methods to detect APTs, which can be categorized into three types based on the way audit logs are utilized: statistical-based methods, rule-based methods, and learning-based methods.
\textbf{Statistical-based methods}, such as Nodoze\cite{hassan2019nodoze}, Swift\cite{hassan2020we}, and PRIOTRACKER\cite{liu2018towards}, quantify the rarity of statistical elements within the provenance graph, assigning suspiciousness scores to these elements, ultimately leading to anomaly detection.
\textbf{Rule-based methods}, like Holmes\cite{milajerdi2019holmes}, SLEUTH\cite{hossain2017sleuth}, and \cite{hassan2020tactical}, collect prior attacks to help security personnel build attack rule libraries, enabling the detection of unknown attacks by matching them to predefined rules.
\textbf{Learning-based methods} utilize machine learning techniques to model benign system behavior or attack patterns in order to identify unknown anomalous behaviors. Relevant work in this category includes \cite{jia2024magic, zengy2022shadewatcher, cheng2023kairos, wang2022threatrace, han2020unicorn, rehman2024flash}.

\noindent \textbf{Knowledge Distillation}. 
Knowledge distillation was first proposed by \cite{hinton2015distilling} and aims to transfer knowledge from a complex model (referred to as the teacher model) to a smaller model (referred to as the student model). This process enables the student model to reduce time and space complexity without compromising prediction quality, thus simplifying model inference and enhancing performance. While knowledge distillation has been widely applied in the field of computer vision, recent studies have extended the basic framework to other domains.
Some works have also focused on the design of the student model. For example, \cite{yang2021extract} was the first to design a student model that is non-GCN-based, while \cite{tian2019contrastive} introduced contrastive learning into the distillation process to improve the student model's ability to learn representations.

%% file: conclusion.tex
\section{Conclusion}
\label{sec:conclusoin}

In this paper, we innovatively propose a novel APTs detection method called \Vodka, which integrates knowledge distillation and provenance-based APTs detection. \Vodka introduces a graph Laplacian-based approach for neighbor denoising and signal smoothing on provenance graphs. Additionally, \Vodka designs a knowledge distillation framework that distills GNNs knowledge from a complex teacher model into a lightweight student model. The student model combines feature transformation and label propagation, allowing for low-cost detection after training and reconstruction of the attack chain. We evaluate \Vodka's performance on three widely used datasets, demonstrating excellent detection results.

%% file: appendix.tex
\appendix \section{Appendix}
\label{sec:appendix}

\subsection{Provenance Graph Description}
Table \ref{tab:node_edge} provides a description of the system event types and the edge relationships in the provenance graph.

\begin{table}[H]
	\renewcommand\arraystretch{1.3}
	\footnotesize
	\centering
	\begin{tabular}{ll}
		\hline
		System event type                              & Relation Description                                             \\ \hline
		Process $\rightarrow$ R1 $\rightarrow$ Process & "R1": "fork", "execute", "exit", "clone", "change",etc. \\
		Process $\rightarrow$ R2 $\rightarrow$ File    & "R2": "read", "open", "close", "write", "rename",etc.   \\
		Process $\rightarrow$ R3 $\rightarrow$ Netflow & "R3": "connect", "send", "recv", "read", "close",etc.   \\
		Process $\rightarrow$ R4 $\rightarrow$ Memory  & "R4": "read", "mprotect", "mmap", etc.                  \\ \hline
	\end{tabular}
	\caption{System event and Rlation description.}
	\label{tab:node_edge}
\end{table}

\subsection{Datasets}
We provide a detailed description of the three datasets used(Table \ref{tab:StreamSpot}, Table \ref{tab:Unicorn}, Table \ref{tab:Darpa-E3}), including their specific contents and original sizes.

\begin{table}[H]
	\renewcommand\arraystretch{1.1}    
	\tabcolsep=5pt
	\normalsize

		\begin{tabular}{cccc}
			\hline
			Scenarios  & \# of Graphs & \begin{tabular}[c]{@{}c@{}}Average \# of \\ Nodes\end{tabular} & \begin{tabular}[c]{@{}c@{}}Average \# of\\ Edges\end{tabular} \\ \hline
			YouTube    & 100          & 8,292                                                          & 113,229                                                        \\
			Gmail      & 100          & 6,826                                                          & 37,382                                                        \\
			Video Game & 100          & 8,636                                                          & 112,958                                                       \\
			Download   & 100          & 8,830                                                          & 310,814                                                        \\
			CNN        & 100          & 8,989                                                          & 294,903                                                       \\
			Attack     & 100          & 8,890                                                          & 28,423                                                        \\ \hline
		\end{tabular}
		\caption{Overview of StreamSpot Dataset.}
		\label{tab:StreamSpot}
\end{table}

\begin{table}[H]
	\renewcommand\arraystretch{1.1}    
	\tabcolsep=3pt

		\begin{tabular}{ccccc}
			\hline
			Scenarios & \begin{tabular}[c]{@{}c@{}}\# Log\\ pieces\end{tabular} & \begin{tabular}[c]{@{}c@{}}Average \# of \\ Entity\end{tabular} & \begin{tabular}[c]{@{}c@{}}Avg \# of\\ Interaction\end{tabular} & \begin{tabular}[c]{@{}c@{}}\# of Raw \\ DataSize(GB)\end{tabular} \\ \hline
			Benign    & 125           & 265,424                                                         & 975,226                                                         & 64.0                                                              \\
			Attack    & 25            & 257,156                                                         & 957,968                                                         & 12.6                                                              \\ \hline
		\end{tabular}
		\caption{Overview of Unicorn Wget Datasets.}
		\label{tab:Unicorn}
\end{table}

\begin{table}[H]
	\renewcommand\arraystretch{1.1}    
	\tabcolsep=3pt

		\begin{tabular}{cccc}
			\hline
			Datasets          & \# of Node & \#of Edges & \begin{tabular}[c]{@{}c@{}}\# of Raw Data \\ Size(GiB)\end{tabular} \\ \hline
			E3-THEIA          & 1,623,966    & 2,874,821    & 17.9                                                               \\
			E3-CADETS         & 1,627,035     & 3,303,264     & 18.3                                                                \\
			E3-TRACE          & 3,288,676    & 4,080,457    & 15.4                                                               \\
			\hline
		\end{tabular}
		\caption{Overview of Darpa-E3 Datasets.}
		\label{tab:Darpa-E3}
\end{table}

\subsection{Initial Parameters for Teacher and Student Models}
We provide more information about the initial settings for both the teacher and student models.

\textbf{Teacher Models} are as follows:
\begin{itemize}[leftmargin=*]
	\item \textbf{GCN}: 2 layers, 64 hidden units, learning rate of 0.01, dropout probability of 0.8, learning rate decay of 0.01.
	\item \textbf{GAT}: 2 layers, 8 attention heads, 64 hidden units, learning rate of 0.01, dropout probability of 0.6, learning rate decay of 0.01, attention dropout probability of 0.3.
	\item \textbf{SAGE}: 128 hidden units, learning rate of 0.01, sample size of 5, batch size of 256, learning rate decay of 0.005.
	\item \textbf{SGC}: 2 layers, learning rate of 0.01, learning rate decay of 0.01.
\end{itemize}

\textbf{Student Model}: The number of layers $K$ is set to 5, the MLP hidden layer has 16 units, dropout rate is 0.2, the Adam optimizer has a learning rate of 0.01, and weight decay is set to 0.001.

\subsection{Interpretability Analysis of the Student Model}

We conduct an interpretability analysis of the student model. Recall that Vodka's student model consists of two main mechanisms: the Feature Transformation (FT) mechanism, powered by MLP, and the Personalized PageRank Random Walk Label Propagation (PRL) mechanism. The label propagation mechanism includes a "jump-back" feature, meaning there is a probability of returning to the previous node, which is controlled by the parameter $\alpha$. Additionally, the final prediction result of the student model is $f_{std}^{k+1}(v)=\beta_{v}f_{FT}(v)+(1-\beta_{v})f_{\mathrm{PRL}}^{k+1}(v)$, and this is controlled by the balancing parameter $\beta$. Our goal is to explore how these two parameters affect the bias of the student model’s prediction results. The experiments were conducted with a GCN as the teacher model and using the DARPA CADETS dataset.

\begin{figure}[H]
	\centering

	\subfigure[Node 1]{
		\includegraphics[width=0.45\columnwidth]{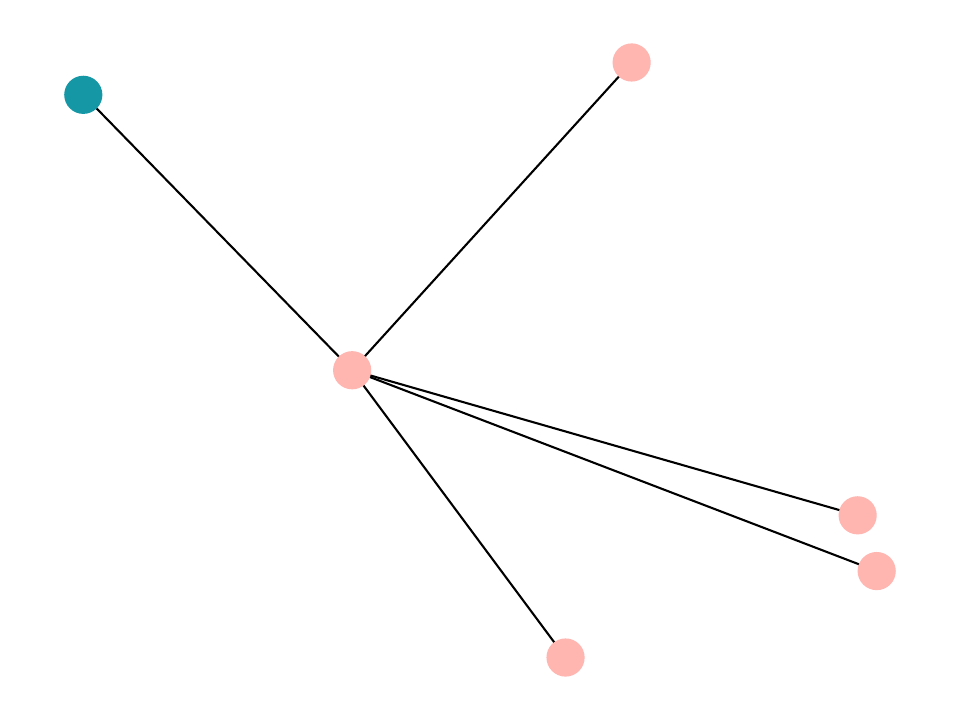}
	}
	\subfigure[Node 2]{
		\includegraphics[width=0.45\columnwidth]{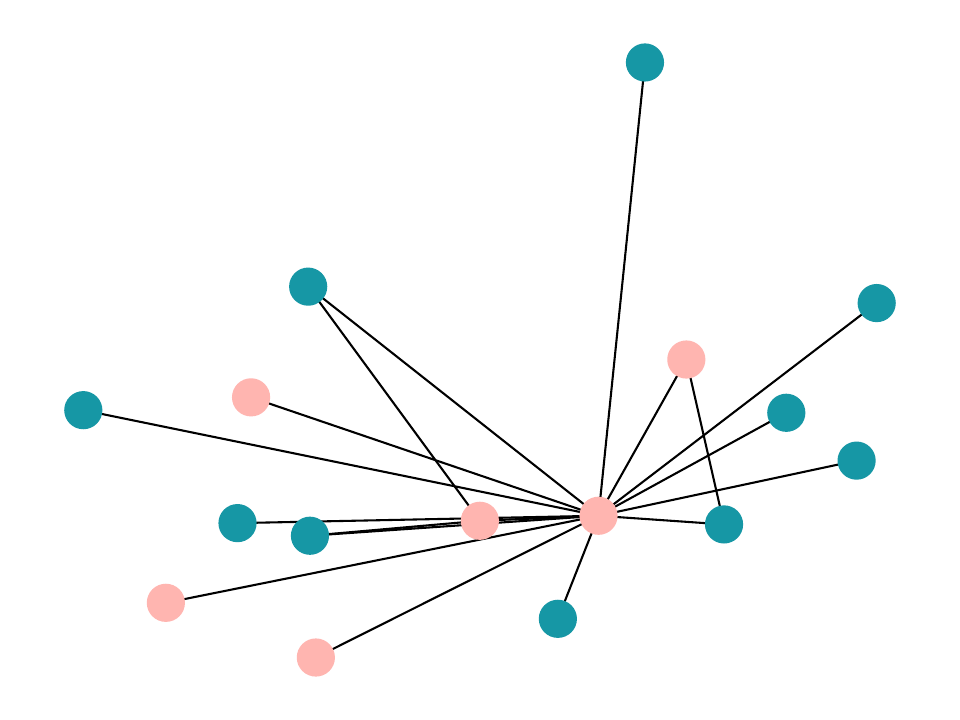}
	}
	\vspace{-0.1in}
	\caption{Study of $\alpha$}
	\label{fig:alpha}
	\vspace{-0.1in}
\end{figure}

\noindent\textbf{Callback Probability $\alpha$}: We dynamically adjust $\alpha$ and plot the neighborhood of the target node when $\alpha$ is at its maximum(Node 1) and minimum(Node 2) values, using different colors to represent labels. As shown in Figure \ref{fig:alpha}, the neighbors of Node 1 are more diverse and numerous compared to those of Node 2. This indicates that nodes with a high $\alpha$ have a higher probability of jumping back to themselves, limiting the diversity of learning from neighboring nodes.

\begin{figure}[H]
	\centering
	
	\subfigure[Node 3]{
		\includegraphics[width=0.45\columnwidth]{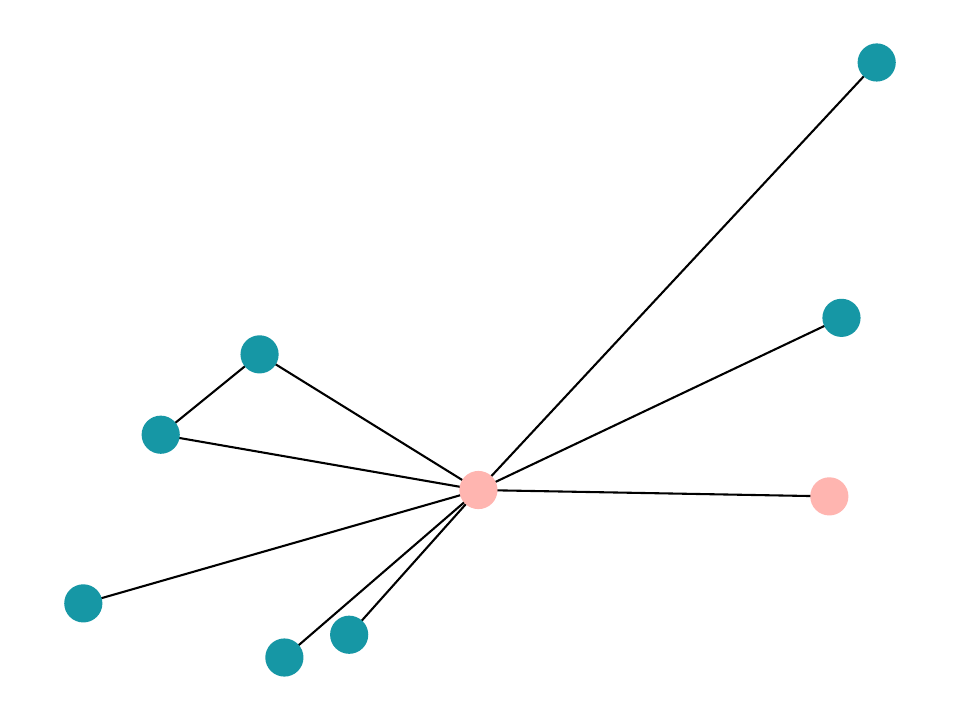}
	}
	\subfigure[Node 4]{
		\includegraphics[width=0.45\columnwidth]{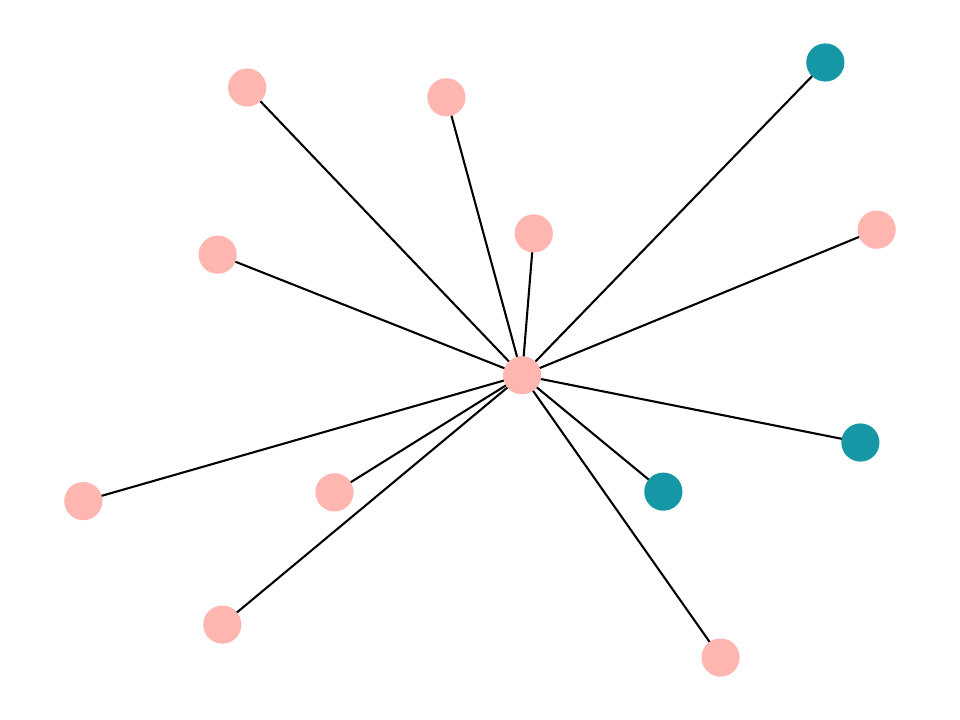}
	}
	\vspace{-0.1in}
	\caption{Study of $\beta$}
	\label{fig:beta}
	\vspace{-0.1in}
\end{figure}

\noindent\textbf{Balancing Parameter $\beta$}: Similarly, we dynamically adjust $\beta$ and plot the neighborhood of the target node when $\beta$ is at its maximum(Node 3) and minimum(Node 4) values, with different colors representing labels. As shown in Figure \ref{fig:beta}, the labels of Node 4’s neighborhood are almost identical, whereas those of Node 3 are almost entirely different. This reflects that the Personalized PageRank Random Walk (PRL) mechanism contributes more to node prediction compared to FT, as when the balancing parameter favors PRL, the prior knowledge from labels and structure plays a more significant role in improving prediction accuracy.

%% file: main.bbl

\begin{thebibliography}{50}


\ifx \showCODEN    \undefined \def \showCODEN     #1{\unskip}     \fi
\ifx \showDOI      \undefined \def \showDOI       #1{#1}\fi
\ifx \showISBNx    \undefined \def \showISBNx     #1{\unskip}     \fi
\ifx \showISBNxiii \undefined \def \showISBNxiii  #1{\unskip}     \fi
\ifx \showISSN     \undefined \def \showISSN      #1{\unskip}     \fi
\ifx \showLCCN     \undefined \def \showLCCN      #1{\unskip}     \fi
\ifx \shownote     \undefined \def \shownote      #1{#1}          \fi
\ifx \showarticletitle \undefined \def \showarticletitle #1{#1}   \fi
\ifx \showURL      \undefined \def \showURL       {\relax}        \fi
\providecommand\bibfield[2]{#2}
\providecommand\bibinfo[2]{#2}
\providecommand\natexlab[1]{#1}
\providecommand\showeprint[2][]{arXiv:#2}

\bibitem[The(2016)]%
        {Thestreamspotdataset}
 \bibinfo{year}{2016}\natexlab{}.
\newblock \bibinfo{title}{The streamspot dataset}.
\newblock \bibinfo{howpublished}{\url{https://github.com/
  sbustreamspot/sbustreamspot-data}}.
\newblock


\bibitem[Equ(2017)]%
        {Equifax}
 \bibinfo{year}{2017}\natexlab{}.
\newblock \bibinfo{title}{Equifax Information Leakage}.
\newblock \bibinfo{howpublished}{\url{https://en.wikipedia.org/wiki/Equifax}}.
\newblock


\bibitem[APT(2020)]%
        {APT42}
 \bibinfo{year}{2020}\natexlab{}.
\newblock \bibinfo{title}{APT42 — Crooked Charms, Cons, and Compromises}.
\newblock
  \bibinfo{howpublished}{\url{https://www.mandiant.com/resources/podcasts/threat-trends-apt42-charms-cons-compromises}}.
\newblock


\bibitem[DAR(2020)]%
        {DARPA}
 \bibinfo{year}{2020}\natexlab{}.
\newblock \bibinfo{title}{Darpa transparent computing program engagement 3 data
  release}.
\newblock \bibinfo{howpublished}{\url{https://github.com/darpa-i2o/
  Transparent-Computing}}.
\newblock


\bibitem[Sol(2020)]%
        {Solarwinds}
 \bibinfo{year}{2020}\natexlab{}.
\newblock \bibinfo{title}{SolaWinds hack}.
\newblock
  \bibinfo{howpublished}{\url{https://en.wikipedia.org/wiki/SolarWinds}}.
\newblock


\bibitem[Alshamrani et~al\mbox{.}(2019)]%
        {alshamrani2019survey}
\bibfield{author}{\bibinfo{person}{Adel Alshamrani}, \bibinfo{person}{Sowmya
  Myneni}, \bibinfo{person}{Ankur Chowdhary}, {and} \bibinfo{person}{Dijiang
  Huang}.} \bibinfo{year}{2019}\natexlab{}.
\newblock \showarticletitle{A survey on advanced persistent threats:
  Techniques, solutions, challenges, and research opportunities}.
\newblock \bibinfo{journal}{\emph{IEEE Communications Surveys \& Tutorials}}
  \bibinfo{volume}{21}, \bibinfo{number}{2} (\bibinfo{year}{2019}),
  \bibinfo{pages}{1851--1877}.
\newblock


\bibitem[Bahmani et~al\mbox{.}(2010)]%
        {bahmani2010fast}
\bibfield{author}{\bibinfo{person}{Bahman Bahmani}, \bibinfo{person}{Abdur
  Chowdhury}, {and} \bibinfo{person}{Ashish Goel}.}
  \bibinfo{year}{2010}\natexlab{}.
\newblock \showarticletitle{Fast incremental and personalized pagerank}.
\newblock \bibinfo{journal}{\emph{arXiv preprint arXiv:1006.2880}}
  (\bibinfo{year}{2010}).
\newblock


\bibitem[Bates et~al\mbox{.}(2015)]%
        {bates2015trustworthy}
\bibfield{author}{\bibinfo{person}{Adam Bates}, \bibinfo{person}{Dave~Jing
  Tian}, \bibinfo{person}{Kevin~RB Butler}, {and} \bibinfo{person}{Thomas
  Moyer}.} \bibinfo{year}{2015}\natexlab{}.
\newblock \showarticletitle{Trustworthy $\{$Whole-System$\}$ provenance for the
  linux kernel}. In \bibinfo{booktitle}{\emph{24th USENIX Security Symposium
  (USENIX Security 15)}}. \bibinfo{pages}{319--334}.
\newblock


\bibitem[Chakraborty et~al\mbox{.}(2021)]%
        {chakraborty2021survey}
\bibfield{author}{\bibinfo{person}{Anirban Chakraborty},
  \bibinfo{person}{Manaar Alam}, \bibinfo{person}{Vishal Dey},
  \bibinfo{person}{Anupam Chattopadhyay}, {and} \bibinfo{person}{Debdeep
  Mukhopadhyay}.} \bibinfo{year}{2021}\natexlab{}.
\newblock \showarticletitle{A survey on adversarial attacks and defences}.
\newblock \bibinfo{journal}{\emph{CAAI Transactions on Intelligence
  Technology}} \bibinfo{volume}{6}, \bibinfo{number}{1} (\bibinfo{year}{2021}),
  \bibinfo{pages}{25--45}.
\newblock


\bibitem[Cheng et~al\mbox{.}(2023)]%
        {cheng2023kairos}
\bibfield{author}{\bibinfo{person}{Zijun Cheng}, \bibinfo{person}{Qiujian Lv},
  \bibinfo{person}{Jinyuan Liang}, \bibinfo{person}{Yan Wang},
  \bibinfo{person}{Degang Sun}, \bibinfo{person}{Thomas Pasquier}, {and}
  \bibinfo{person}{Xueyuan Han}.} \bibinfo{year}{2023}\natexlab{}.
\newblock \showarticletitle{Kairos:: Practical Intrusion Detection and
  Investigation using Whole-system Provenance}.
\newblock \bibinfo{journal}{\emph{arXiv preprint arXiv:2308.05034}}
  (\bibinfo{year}{2023}).
\newblock


\bibitem[Du et~al\mbox{.}(2017)]%
        {du2017deeplog}
\bibfield{author}{\bibinfo{person}{Min Du}, \bibinfo{person}{Feifei Li},
  \bibinfo{person}{Guineng Zheng}, {and} \bibinfo{person}{Vivek Srikumar}.}
  \bibinfo{year}{2017}\natexlab{}.
\newblock \showarticletitle{Deeplog: Anomaly detection and diagnosis from
  system logs through deep learning}. In \bibinfo{booktitle}{\emph{Proceedings
  of the 2017 ACM SIGSAC conference on computer and communications security}}.
  \bibinfo{pages}{1285--1298}.
\newblock


\bibitem[Edler et~al\mbox{.}(2017)]%
        {edler2017infomap}
\bibfield{author}{\bibinfo{person}{Daniel Edler}, \bibinfo{person}{Tha{\'\i}s
  Guedes}, \bibinfo{person}{Alexander Zizka}, \bibinfo{person}{Martin Rosvall},
  {and} \bibinfo{person}{Alexandre Antonelli}.}
  \bibinfo{year}{2017}\natexlab{}.
\newblock \showarticletitle{Infomap bioregions: interactive mapping of
  biogeographical regions from species distributions}.
\newblock \bibinfo{journal}{\emph{Systematic biology}} \bibinfo{volume}{66},
  \bibinfo{number}{2} (\bibinfo{year}{2017}), \bibinfo{pages}{197--204}.
\newblock


\bibitem[Goyal et~al\mbox{.}(2023)]%
        {goyal2023sometimes}
\bibfield{author}{\bibinfo{person}{Akul Goyal}, \bibinfo{person}{Xueyuan Han},
  \bibinfo{person}{Gang Wang}, {and} \bibinfo{person}{Adam Bates}.}
  \bibinfo{year}{2023}\natexlab{}.
\newblock \showarticletitle{Sometimes, you aren’t what you do: Mimicry
  attacks against provenance graph host intrusion detection systems}. In
  \bibinfo{booktitle}{\emph{30th Network and Distributed System Security
  Symposium}}.
\newblock


\bibitem[Hamilton et~al\mbox{.}(2017)]%
        {hamilton2017inductive}
\bibfield{author}{\bibinfo{person}{Will Hamilton}, \bibinfo{person}{Zhitao
  Ying}, {and} \bibinfo{person}{Jure Leskovec}.}
  \bibinfo{year}{2017}\natexlab{}.
\newblock \showarticletitle{Inductive representation learning on large graphs}.
\newblock \bibinfo{journal}{\emph{Advances in neural information processing
  systems}}  \bibinfo{volume}{30} (\bibinfo{year}{2017}).
\newblock


\bibitem[Han et~al\mbox{.}(2020)]%
        {han2020unicorn}
\bibfield{author}{\bibinfo{person}{Xueyuan Han}, \bibinfo{person}{Thomas
  Pasquier}, \bibinfo{person}{Adam Bates}, \bibinfo{person}{James Mickens},
  {and} \bibinfo{person}{Margo Seltzer}.} \bibinfo{year}{2020}\natexlab{}.
\newblock \showarticletitle{Unicorn: Runtime provenance-based detector for
  advanced persistent threats}.
\newblock \bibinfo{journal}{\emph{arXiv preprint arXiv:2001.01525}}
  (\bibinfo{year}{2020}).
\newblock


\bibitem[Hassan et~al\mbox{.}(2020a)]%
        {hassan2020tactical}
\bibfield{author}{\bibinfo{person}{Wajih~Ul Hassan}, \bibinfo{person}{Adam
  Bates}, {and} \bibinfo{person}{Daniel Marino}.}
  \bibinfo{year}{2020}\natexlab{a}.
\newblock \showarticletitle{Tactical provenance analysis for endpoint detection
  and response systems}. In \bibinfo{booktitle}{\emph{2020 IEEE Symposium on
  Security and Privacy (SP)}}. IEEE, \bibinfo{pages}{1172--1189}.
\newblock


\bibitem[Hassan et~al\mbox{.}(2019)]%
        {hassan2019nodoze}
\bibfield{author}{\bibinfo{person}{Wajih~Ul Hassan}, \bibinfo{person}{Shengjian
  Guo}, \bibinfo{person}{Ding Li}, \bibinfo{person}{Zhengzhang Chen},
  \bibinfo{person}{Kangkook Jee}, \bibinfo{person}{Zhichun Li}, {and}
  \bibinfo{person}{Adam Bates}.} \bibinfo{year}{2019}\natexlab{}.
\newblock \showarticletitle{Nodoze: Combatting threat alert fatigue with
  automated provenance triage}. In \bibinfo{booktitle}{\emph{network and
  distributed systems security symposium}}.
\newblock


\bibitem[Hassan et~al\mbox{.}(2020b)]%
        {hassan2020we}
\bibfield{author}{\bibinfo{person}{Wajih~Ul Hassan}, \bibinfo{person}{Ding Li},
  \bibinfo{person}{Kangkook Jee}, \bibinfo{person}{Xiao Yu},
  \bibinfo{person}{Kexuan Zou}, \bibinfo{person}{Dawei Wang},
  \bibinfo{person}{Zhengzhang Chen}, \bibinfo{person}{Zhichun Li},
  \bibinfo{person}{Junghwan Rhee}, \bibinfo{person}{Jiaping Gui},
  {et~al\mbox{.}}} \bibinfo{year}{2020}\natexlab{b}.
\newblock \showarticletitle{This is why we can’t cache nice things:
  Lightning-fast threat hunting using suspicion-based hierarchical storage}. In
  \bibinfo{booktitle}{\emph{Proceedings of the 36th Annual Computer Security
  Applications Conference}}. \bibinfo{pages}{165--178}.
\newblock


\bibitem[Hinton et~al\mbox{.}(2015)]%
        {hinton2015distilling}
\bibfield{author}{\bibinfo{person}{Geoffrey Hinton}, \bibinfo{person}{Oriol
  Vinyals}, {and} \bibinfo{person}{Jeff Dean}.}
  \bibinfo{year}{2015}\natexlab{}.
\newblock \showarticletitle{Distilling the Knowledge in a Neural Network}.
\newblock \bibinfo{journal}{\emph{stat}}  \bibinfo{volume}{1050}
  (\bibinfo{year}{2015}), \bibinfo{pages}{9}.
\newblock


\bibitem[Hossain et~al\mbox{.}(2017)]%
        {hossain2017sleuth}
\bibfield{author}{\bibinfo{person}{Md~Nahid Hossain}, \bibinfo{person}{Sadegh~M
  Milajerdi}, \bibinfo{person}{Junao Wang}, \bibinfo{person}{Birhanu Eshete},
  \bibinfo{person}{Rigel Gjomemo}, \bibinfo{person}{R Sekar},
  \bibinfo{person}{Scott Stoller}, {and} \bibinfo{person}{VN Venkatakrishnan}.}
  \bibinfo{year}{2017}\natexlab{}.
\newblock \showarticletitle{SLEUTH: Real-time attack scenario reconstruction
  from COTS audit data}. In \bibinfo{booktitle}{\emph{26th USENIX Security
  Symposium (USENIX Security 17)}}. \bibinfo{pages}{487--504}.
\newblock


\bibitem[Iscen et~al\mbox{.}(2019)]%
        {iscen2019label}
\bibfield{author}{\bibinfo{person}{Ahmet Iscen}, \bibinfo{person}{Giorgos
  Tolias}, \bibinfo{person}{Yannis Avrithis}, {and} \bibinfo{person}{Ondrej
  Chum}.} \bibinfo{year}{2019}\natexlab{}.
\newblock \showarticletitle{Label propagation for deep semi-supervised
  learning}. In \bibinfo{booktitle}{\emph{Proceedings of the IEEE/CVF
  conference on computer vision and pattern recognition}}.
  \bibinfo{pages}{5070--5079}.
\newblock


\bibitem[Jacob et~al\mbox{.}(2008)]%
        {jacob2008systemtap}
\bibfield{author}{\bibinfo{person}{Bart Jacob}, \bibinfo{person}{Paul Larson},
  \bibinfo{person}{B Leitao}, {and} \bibinfo{person}{SAMM Da~Silva}.}
  \bibinfo{year}{2008}\natexlab{}.
\newblock \showarticletitle{SystemTap: instrumenting the Linux kernel for
  analyzing performance and functional problems}.
\newblock \bibinfo{journal}{\emph{IBM Redbook}}  \bibinfo{volume}{116}
  (\bibinfo{year}{2008}).
\newblock


\bibitem[Jia et~al\mbox{.}(2024)]%
        {jia2024magic}
\bibfield{author}{\bibinfo{person}{Zian Jia}, \bibinfo{person}{Yun Xiong},
  \bibinfo{person}{Yuhong Nan}, \bibinfo{person}{Yao Zhang},
  \bibinfo{person}{Jinjing Zhao}, {and} \bibinfo{person}{Mi Wen}.}
  \bibinfo{year}{2024}\natexlab{}.
\newblock \showarticletitle{$\{$MAGIC$\}$: Detecting Advanced Persistent
  Threats via Masked Graph Representation Learning}. In
  \bibinfo{booktitle}{\emph{33rd USENIX Security Symposium (USENIX Security
  24)}}. \bibinfo{pages}{5197--5214}.
\newblock


\bibitem[Kapoor et~al\mbox{.}(2021)]%
        {kapoor2021prov}
\bibfield{author}{\bibinfo{person}{Maya Kapoor}, \bibinfo{person}{Joshua
  Melton}, \bibinfo{person}{Michael Ridenhour}, \bibinfo{person}{Siddharth
  Krishnan}, {and} \bibinfo{person}{Thomas Moyer}.}
  \bibinfo{year}{2021}\natexlab{}.
\newblock \showarticletitle{PROV-GEM: automated provenance analysis framework
  using graph embeddings}. In \bibinfo{booktitle}{\emph{2021 20th IEEE
  International Conference on Machine Learning and Applications (ICMLA)}}.
  IEEE, \bibinfo{pages}{1720--1727}.
\newblock


\bibitem[Khaleefa and Abdulah(2022)]%
        {khaleefa2022concept}
\bibfield{author}{\bibinfo{person}{Eman~J Khaleefa} {and}
  \bibinfo{person}{Dhahair~A Abdulah}.} \bibinfo{year}{2022}\natexlab{}.
\newblock \showarticletitle{Concept and difficulties of advanced persistent
  threats (APT): Survey}.
\newblock \bibinfo{journal}{\emph{International Journal of Nonlinear Analysis
  and Applications}} \bibinfo{volume}{13}, \bibinfo{number}{1}
  (\bibinfo{year}{2022}), \bibinfo{pages}{4037--4052}.
\newblock


\bibitem[Kipf and Welling(2016)]%
        {kipf2016semi}
\bibfield{author}{\bibinfo{person}{Thomas~N Kipf} {and} \bibinfo{person}{Max
  Welling}.} \bibinfo{year}{2016}\natexlab{}.
\newblock \showarticletitle{Semi-supervised classification with graph
  convolutional networks}.
\newblock \bibinfo{journal}{\emph{arXiv preprint arXiv:1609.02907}}
  (\bibinfo{year}{2016}).
\newblock


\bibitem[Li et~al\mbox{.}(2019)]%
        {li2019label}
\bibfield{author}{\bibinfo{person}{Qimai Li}, \bibinfo{person}{Xiao-Ming Wu},
  \bibinfo{person}{Han Liu}, \bibinfo{person}{Xiaotong Zhang}, {and}
  \bibinfo{person}{Zhichao Guan}.} \bibinfo{year}{2019}\natexlab{}.
\newblock \showarticletitle{Label efficient semi-supervised learning via graph
  filtering}. In \bibinfo{booktitle}{\emph{Proceedings of the IEEE/CVF
  conference on computer vision and pattern recognition}}.
  \bibinfo{pages}{9582--9591}.
\newblock


\bibitem[Li et~al\mbox{.}(2021)]%
        {li2021threat}
\bibfield{author}{\bibinfo{person}{Zhenyuan Li}, \bibinfo{person}{Qi~Alfred
  Chen}, \bibinfo{person}{Runqing Yang}, \bibinfo{person}{Yan Chen}, {and}
  \bibinfo{person}{Wei Ruan}.} \bibinfo{year}{2021}\natexlab{}.
\newblock \showarticletitle{Threat detection and investigation with
  system-level provenance graphs: A survey}.
\newblock \bibinfo{journal}{\emph{Computers \& Security}}
  \bibinfo{volume}{106} (\bibinfo{year}{2021}), \bibinfo{pages}{102282}.
\newblock


\bibitem[Liu et~al\mbox{.}(2019)]%
        {liu2019log2vec}
\bibfield{author}{\bibinfo{person}{Fucheng Liu}, \bibinfo{person}{Yu Wen},
  \bibinfo{person}{Dongxue Zhang}, \bibinfo{person}{Xihe Jiang},
  \bibinfo{person}{Xinyu Xing}, {and} \bibinfo{person}{Dan Meng}.}
  \bibinfo{year}{2019}\natexlab{}.
\newblock \showarticletitle{Log2vec: A heterogeneous graph embedding based
  approach for detecting cyber threats within enterprise}. In
  \bibinfo{booktitle}{\emph{Proceedings of the 2019 ACM SIGSAC conference on
  computer and communications security}}. \bibinfo{pages}{1777--1794}.
\newblock


\bibitem[Liu et~al\mbox{.}(2018)]%
        {liu2018towards}
\bibfield{author}{\bibinfo{person}{Yushan Liu}, \bibinfo{person}{Mu Zhang},
  \bibinfo{person}{Ding Li}, \bibinfo{person}{Kangkook Jee},
  \bibinfo{person}{Zhichun Li}, \bibinfo{person}{Zhenyu Wu},
  \bibinfo{person}{Junghwan Rhee}, {and} \bibinfo{person}{Prateek Mittal}.}
  \bibinfo{year}{2018}\natexlab{}.
\newblock \showarticletitle{Towards a Timely Causality Analysis for Enterprise
  Security.}. In \bibinfo{booktitle}{\emph{NDSS}}.
\newblock


\bibitem[Ma and Zhang(2015)]%
        {ma2015using}
\bibfield{author}{\bibinfo{person}{Long Ma} {and} \bibinfo{person}{Yanqing
  Zhang}.} \bibinfo{year}{2015}\natexlab{}.
\newblock \showarticletitle{Using Word2Vec to process big text data}. In
  \bibinfo{booktitle}{\emph{2015 IEEE International Conference on Big Data (Big
  Data)}}. IEEE, \bibinfo{pages}{2895--2897}.
\newblock


\bibitem[Milajerdi et~al\mbox{.}(2019)]%
        {milajerdi2019holmes}
\bibfield{author}{\bibinfo{person}{Sadegh~M Milajerdi}, \bibinfo{person}{Rigel
  Gjomemo}, \bibinfo{person}{Birhanu Eshete}, \bibinfo{person}{Ramachandran
  Sekar}, {and} \bibinfo{person}{VN Venkatakrishnan}.}
  \bibinfo{year}{2019}\natexlab{}.
\newblock \showarticletitle{Holmes: real-time apt detection through correlation
  of suspicious information flows}. In \bibinfo{booktitle}{\emph{2019 IEEE
  Symposium on Security and Privacy (SP)}}. IEEE, \bibinfo{pages}{1137--1152}.
\newblock


\bibitem[Ongun et~al\mbox{.}(2021)]%
        {ongun2021living}
\bibfield{author}{\bibinfo{person}{Talha Ongun}, \bibinfo{person}{Jack~W
  Stokes}, \bibinfo{person}{Jonathan~Bar Or}, \bibinfo{person}{Ke Tian},
  \bibinfo{person}{Farid Tajaddodianfar}, \bibinfo{person}{Joshua Neil},
  \bibinfo{person}{Christian Seifert}, \bibinfo{person}{Alina Oprea}, {and}
  \bibinfo{person}{John~C Platt}.} \bibinfo{year}{2021}\natexlab{}.
\newblock \showarticletitle{Living-off-the-land command detection using active
  learning}. In \bibinfo{booktitle}{\emph{Proceedings of the 24th International
  Symposium on Research in Attacks, Intrusions and Defenses}}.
  \bibinfo{pages}{442--455}.
\newblock


\bibitem[Paccagnella et~al\mbox{.}(2020)]%
        {paccagnella2020custos}
\bibfield{author}{\bibinfo{person}{Riccardo Paccagnella},
  \bibinfo{person}{Pubali Datta}, \bibinfo{person}{Wajih~Ul Hassan},
  \bibinfo{person}{Adam Bates}, \bibinfo{person}{Christopher Fletcher},
  \bibinfo{person}{Andrew Miller}, {and} \bibinfo{person}{Dave Tian}.}
  \bibinfo{year}{2020}\natexlab{}.
\newblock \showarticletitle{Custos: Practical tamper-evident auditing of
  operating systems using trusted execution}. In
  \bibinfo{booktitle}{\emph{Network and distributed system security
  symposium}}.
\newblock


\bibitem[Pasquier et~al\mbox{.}(2017)]%
        {pasquier2017practical}
\bibfield{author}{\bibinfo{person}{Thomas Pasquier}, \bibinfo{person}{Xueyuan
  Han}, \bibinfo{person}{Mark Goldstein}, \bibinfo{person}{Thomas Moyer},
  \bibinfo{person}{David Eyers}, \bibinfo{person}{Margo Seltzer}, {and}
  \bibinfo{person}{Jean Bacon}.} \bibinfo{year}{2017}\natexlab{}.
\newblock \showarticletitle{Practical whole-system provenance capture}. In
  \bibinfo{booktitle}{\emph{Proceedings of the 2017 Symposium on Cloud
  Computing}}. \bibinfo{pages}{405--418}.
\newblock


\bibitem[Paszke et~al\mbox{.}(2019)]%
        {paszke2019pytorch}
\bibfield{author}{\bibinfo{person}{Adam Paszke}, \bibinfo{person}{Sam Gross},
  \bibinfo{person}{Francisco Massa}, \bibinfo{person}{Adam Lerer},
  \bibinfo{person}{James Bradbury}, \bibinfo{person}{Gregory Chanan},
  \bibinfo{person}{Trevor Killeen}, \bibinfo{person}{Zeming Lin},
  \bibinfo{person}{Natalia Gimelshein}, \bibinfo{person}{Luca Antiga},
  {et~al\mbox{.}}} \bibinfo{year}{2019}\natexlab{}.
\newblock \showarticletitle{Pytorch: An imperative style, high-performance deep
  learning library}.
\newblock \bibinfo{journal}{\emph{Advances in neural information processing
  systems}}  \bibinfo{volume}{32} (\bibinfo{year}{2019}).
\newblock


\bibitem[Rehman et~al\mbox{.}(2024)]%
        {rehman2024flash}
\bibfield{author}{\bibinfo{person}{Mati~Ur Rehman}, \bibinfo{person}{Hadi
  Ahmadi}, {and} \bibinfo{person}{Wajih~Ul Hassan}.}
  \bibinfo{year}{2024}\natexlab{}.
\newblock \showarticletitle{FLASH: A Comprehensive Approach to Intrusion
  Detection via Provenance Graph Representation Learning}. In
  \bibinfo{booktitle}{\emph{2024 IEEE Symposium on Security and Privacy (SP)}}.
  IEEE Computer Society, \bibinfo{pages}{139--139}.
\newblock


\bibitem[Sharma et~al\mbox{.}(2023)]%
        {sharma2023advanced}
\bibfield{author}{\bibinfo{person}{Amit Sharma}, \bibinfo{person}{Brij~B
  Gupta}, \bibinfo{person}{Awadhesh~Kumar Singh}, {and} \bibinfo{person}{VK
  Saraswat}.} \bibinfo{year}{2023}\natexlab{}.
\newblock \showarticletitle{Advanced Persistent Threats (APT): evolution,
  anatomy, attribution and countermeasures}.
\newblock \bibinfo{journal}{\emph{Journal of Ambient Intelligence and Humanized
  Computing}} \bibinfo{volume}{14}, \bibinfo{number}{7} (\bibinfo{year}{2023}),
  \bibinfo{pages}{9355--9381}.
\newblock


\bibitem[Tang et~al\mbox{.}(2015)]%
        {tang2015extreme}
\bibfield{author}{\bibinfo{person}{Jiexiong Tang}, \bibinfo{person}{Chenwei
  Deng}, {and} \bibinfo{person}{Guang-Bin Huang}.}
  \bibinfo{year}{2015}\natexlab{}.
\newblock \showarticletitle{Extreme learning machine for multilayer
  perceptron}.
\newblock \bibinfo{journal}{\emph{IEEE transactions on neural networks and
  learning systems}} \bibinfo{volume}{27}, \bibinfo{number}{4}
  (\bibinfo{year}{2015}), \bibinfo{pages}{809--821}.
\newblock


\bibitem[Tian et~al\mbox{.}(2019)]%
        {tian2019contrastive}
\bibfield{author}{\bibinfo{person}{Yonglong Tian}, \bibinfo{person}{Dilip
  Krishnan}, {and} \bibinfo{person}{Phillip Isola}.}
  \bibinfo{year}{2019}\natexlab{}.
\newblock \showarticletitle{Contrastive representation distillation}.
\newblock \bibinfo{journal}{\emph{arXiv preprint arXiv:1910.10699}}
  (\bibinfo{year}{2019}).
\newblock


\bibitem[Tramer and Boneh(2019)]%
        {tramer2019adversarial}
\bibfield{author}{\bibinfo{person}{Florian Tramer} {and} \bibinfo{person}{Dan
  Boneh}.} \bibinfo{year}{2019}\natexlab{}.
\newblock \showarticletitle{Adversarial training and robustness for multiple
  perturbations}.
\newblock \bibinfo{journal}{\emph{Advances in neural information processing
  systems}}  \bibinfo{volume}{32} (\bibinfo{year}{2019}).
\newblock


\bibitem[Veli{\v{c}}kovi{\'c} et~al\mbox{.}(2017)]%
        {velivckovic2017graph}
\bibfield{author}{\bibinfo{person}{Petar Veli{\v{c}}kovi{\'c}},
  \bibinfo{person}{Guillem Cucurull}, \bibinfo{person}{Arantxa Casanova},
  \bibinfo{person}{Adriana Romero}, \bibinfo{person}{Pietro Lio}, {and}
  \bibinfo{person}{Yoshua Bengio}.} \bibinfo{year}{2017}\natexlab{}.
\newblock \showarticletitle{Graph attention networks}.
\newblock \bibinfo{journal}{\emph{arXiv preprint arXiv:1710.10903}}
  (\bibinfo{year}{2017}).
\newblock


\bibitem[Wang and Leskovec(2020)]%
        {wang2020unifying}
\bibfield{author}{\bibinfo{person}{Hongwei Wang} {and} \bibinfo{person}{Jure
  Leskovec}.} \bibinfo{year}{2020}\natexlab{}.
\newblock \showarticletitle{Unifying graph convolutional neural networks and
  label propagation}.
\newblock \bibinfo{journal}{\emph{arXiv preprint arXiv:2002.06755}}
  (\bibinfo{year}{2020}).
\newblock


\bibitem[Wang(2019)]%
        {wang2019deep}
\bibfield{author}{\bibinfo{person}{Minjie~Yu Wang}.}
  \bibinfo{year}{2019}\natexlab{}.
\newblock \showarticletitle{Deep graph library: Towards efficient and scalable
  deep learning on graphs}. In \bibinfo{booktitle}{\emph{ICLR workshop on
  representation learning on graphs and manifolds}}.
\newblock


\bibitem[Wang et~al\mbox{.}(2022)]%
        {wang2022threatrace}
\bibfield{author}{\bibinfo{person}{Su Wang}, \bibinfo{person}{Zhiliang Wang},
  \bibinfo{person}{Tao Zhou}, \bibinfo{person}{Hongbin Sun},
  \bibinfo{person}{Xia Yin}, \bibinfo{person}{Dongqi Han}, \bibinfo{person}{Han
  Zhang}, \bibinfo{person}{Xingang Shi}, {and} \bibinfo{person}{Jiahai Yang}.}
  \bibinfo{year}{2022}\natexlab{}.
\newblock \showarticletitle{Threatrace: Detecting and tracing host-based
  threats in node level through provenance graph learning}.
\newblock \bibinfo{journal}{\emph{IEEE Transactions on Information Forensics
  and Security}}  \bibinfo{volume}{17} (\bibinfo{year}{2022}),
  \bibinfo{pages}{3972--3987}.
\newblock


\bibitem[Wu et~al\mbox{.}(2019)]%
        {wu2019simplifying}
\bibfield{author}{\bibinfo{person}{Felix Wu}, \bibinfo{person}{Amauri Souza},
  \bibinfo{person}{Tianyi Zhang}, \bibinfo{person}{Christopher Fifty},
  \bibinfo{person}{Tao Yu}, {and} \bibinfo{person}{Kilian Weinberger}.}
  \bibinfo{year}{2019}\natexlab{}.
\newblock \showarticletitle{Simplifying graph convolutional networks}. In
  \bibinfo{booktitle}{\emph{International conference on machine learning}}.
  PMLR, \bibinfo{pages}{6861--6871}.
\newblock


\bibitem[Xia et~al\mbox{.}(2019)]%
        {xia2019random}
\bibfield{author}{\bibinfo{person}{Feng Xia}, \bibinfo{person}{Jiaying Liu},
  \bibinfo{person}{Hansong Nie}, \bibinfo{person}{Yonghao Fu},
  \bibinfo{person}{Liangtian Wan}, {and} \bibinfo{person}{Xiangjie Kong}.}
  \bibinfo{year}{2019}\natexlab{}.
\newblock \showarticletitle{Random walks: A review of algorithms and
  applications}.
\newblock \bibinfo{journal}{\emph{IEEE Transactions on Emerging Topics in
  Computational Intelligence}} \bibinfo{volume}{4}, \bibinfo{number}{2}
  (\bibinfo{year}{2019}), \bibinfo{pages}{95--107}.
\newblock


\bibitem[Yang et~al\mbox{.}(2021)]%
        {yang2021extract}
\bibfield{author}{\bibinfo{person}{Cheng Yang}, \bibinfo{person}{Jiawei Liu},
  {and} \bibinfo{person}{Chuan Shi}.} \bibinfo{year}{2021}\natexlab{}.
\newblock \showarticletitle{Extract the knowledge of graph neural networks and
  go beyond it: An effective knowledge distillation framework}. In
  \bibinfo{booktitle}{\emph{Proceedings of the web conference 2021}}.
  \bibinfo{pages}{1227--1237}.
\newblock


\bibitem[Yu et~al\mbox{.}(2019)]%
        {yu2019review}
\bibfield{author}{\bibinfo{person}{Yong Yu}, \bibinfo{person}{Xiaosheng Si},
  \bibinfo{person}{Changhua Hu}, {and} \bibinfo{person}{Jianxun Zhang}.}
  \bibinfo{year}{2019}\natexlab{}.
\newblock \showarticletitle{A review of recurrent neural networks: LSTM cells
  and network architectures}.
\newblock \bibinfo{journal}{\emph{Neural computation}} \bibinfo{volume}{31},
  \bibinfo{number}{7} (\bibinfo{year}{2019}), \bibinfo{pages}{1235--1270}.
\newblock


\bibitem[Zengy et~al\mbox{.}(2022)]%
        {zengy2022shadewatcher}
\bibfield{author}{\bibinfo{person}{Jun Zengy}, \bibinfo{person}{Xiang Wang},
  \bibinfo{person}{Jiahao Liu}, \bibinfo{person}{Yinfang Chen},
  \bibinfo{person}{Zhenkai Liang}, \bibinfo{person}{Tat-Seng Chua}, {and}
  \bibinfo{person}{Zheng~Leong Chua}.} \bibinfo{year}{2022}\natexlab{}.
\newblock \showarticletitle{Shadewatcher: Recommendation-guided cyber threat
  analysis using system audit records}. In \bibinfo{booktitle}{\emph{2022 IEEE
  Symposium on Security and Privacy (SP)}}. IEEE, \bibinfo{pages}{489--506}.
\newblock


\end{thebibliography}
